\documentstyle[
aps,
preprint,
epsfig,
bm,
tighten,
]{revtex}

\newcommand{\beq}{\begin{equation}}
\newcommand{\eeq}{\end{equation}}
\newcommand{\beqa}{\begin{eqnarray}}
\newcommand{\eeqa}{\end{eqnarray}}
\newcommand{\Psib}{\overline{\Psi}}

\newcommand{\Lcal}{{\mathcal L}}
\newcommand{\p}{\partial}
\newcommand{\se}{\Sigma}
\begin{document}
\draft


\title{Density dependent hadron field theory for asymmetric nuclear
matter and exotic nuclei}
\author{F. Hofmann, C. M. Keil, H. Lenske}
\address{Institut f\"ur Theoretische Physik, Universit\"at Gie\ss en,
         Heinrich-Buff-Ring 16, 35392 Gie\ss en, Germany}

\date{\today}

\maketitle

\begin{abstract}

The density dependent relativistic hadron field (DDRH) theory is
applied to strongly asymmetric nuclear matter and finite nuclei far
off stability. A new set of in-medium meson-nucleon vertices is
derived from Dirac-Brueckner Hartree-Fock (DBHF) calculations in
asymmetric matter, now accounting also for the density dependence of
isovector coupling constants. The scalar-isovector $\delta$ meson is
included. Nuclear matter calculations show that it is necessary to
introduce a momentum correction in the extraction of coupling
constants from the DBHF self-energies in order to reproduce the DBHF
equation of state by DDRH mean-field calculations. The properties of
DDRH vertices derived from the Groningen and the Bonn A 
nucleon-nucleon (NN) potentials 
are compared in nuclear matter calculations
and for finite nuclei. Relativistic Hartree results for binding
energies, charge radii, separation energies and shell gaps for the Ni
and Sn isotopic chains are presented. Using the momentum corrected
vertices an overall agreement to data on a level of a few percent is
obtained. In the accessible range of asymmetries the $\delta$ meson
contributions to the self-energies are found to be of minor importance
but asymmetry dependent fluctuations may occur.

\end{abstract}



\section{Introduction}
\label{sec:Intro}

The modern approach to nuclear structure is based on relativistic
models describing nuclear matter and finite nuclei as a strongly
interacting systems of baryons and mesons. Starting from a
Lagrangian formulation a phenomenological hadronic field theory
is obtained by adjusting the meson-nucleon coupling constants to
various properties of infinite nuclear matter and finite nuclei
\cite{Reinhard:86,Gambhir:90}. A connection to free space
nucleon-nucleon interactions is not attempted. The prototype for
such an approach is relativistic mean-field theory (RMF)
\cite{Walecka:74,Serot:86} where nuclear forces are obtained from
the virtual exchange of mesons, finally leading to condensed
classically fields produced by nucleonic sources. Using this
procedure recent RMF models have been remarkably successful in
describing nuclei over the entire range of the periodic table
\cite{Reinhard:88,Sharma:93,Sugahara:94,Lalazissis:97,Sharma:00}.
In order to improve results cubic and quartic self-interactions
of the mesons fields had to be introduced
\cite{Boguta:77,Bodmer:91}. Although higher order scalar
self-interactions can be motivated by vacuum renormalization
\cite{Serot:86} in practice the strengths of the self-couplings
are determined in a purely phenomenological way. In mean-field
approximation the mesonic self-interactions correspond
effectively to higher order density dependent contributions.
Using up to quartic terms a good description of nuclear matter
and finite nuclei is obtained but, depending on the sign of
especially the quartic scalar $\sigma$ self-interaction,
instabilities occur in the region above saturation density
\cite{Reinhard:88}.

A more fundamental - but also more elaborate - approach is to
derive in-medium interactions microscopically. An appropriate and
successful method is Dirac-Brueckner theory (DB). Using realistic
NN potentials in-medium interactions are derived by a complete
re-summation of (two-body) ladder diagrams. A break-through was
obtained with relativistic Brueckner theory which reproduces the
empirical saturation properties of nuclear matter reasonably well
\cite{Anastasio:83,Horowitz:87,TerHaar:87,Brockmann:90,Boersma:94,Huber:95,deJong:98a}.
Since full-scale DB calculations for finite nuclei are not
feasible a practical approach is to apply infinite matter DB
results in local density approximation (LDA) to finite nuclei
\cite{Brockmann:92,Haddad:93,Boersma:94b}. Retaining a Lagrangian
formulation this is achieved by introducing density dependent
meson-nucleon coupling constants taken from DB self-energies
\cite{Brockmann:92}. In \cite{Lenske:95,Fuchs:95} it was pointed
out that such an approach does not comply with relativity and
thermodynamics. A fully covariant and thermodynamically
consistent field theory, however, is obtained by treating the
interaction vertices on the level of the Lagrangian as
Lorentz-scalar functionals of the field operators. In the density
dependent relativistic hadron field (DDRH) theory
\cite{Lenske:95,Fuchs:95} the medium dependence of the vertices
is expressed by functionals of the baryon field operators. An
important difference to the RMF treatment of non-linearities is
that the DDRH approach accounts for quantal fluctuations of the
baryon fields even in the ground state. Such effects contribute
as rearrangement self-energies to the baryon field equations
describing the static polarization of the background medium by a
nucleon \cite{Fuchs:95,Negele:82}. Since DDRH theory provides a
systematic expansion of interactions in terms of higher order
baryon-baryon correlation functions \cite{Lenske:95,Fuchs:95},
extensions beyond the mean-level are in principle possible.

In mean-field approximation DDRH theory reduces to a Hartree
description with density dependent coupling constants similar to
the initial proposal of Brockmann and Toki \cite{Brockmann:92}.
The rearrangement contributions significantly improve the binding
energies and radii of finite nuclei. Several calculations in the
DDRH model for stable nuclei have been performed
\cite{Fuchs:95,Ineichen:95,Shen:97,Cescato:98} using density
dependent vertices derived from DB calculations with the Bonn A
potential \cite{Brockmann:90,Machleidt:89}. Recently, a
phenomenological approach to DDRH theory was presented
\cite{Typel:99} by determining the density dependence of the
vertices empirically. Descriptions of finite nuclei of a quality
comparable to non-linear RMF models were obtained. Extensions to
the baryon octet sector and hypernuclei are discussed in
\cite{Keil:00}.

The main intention of this paper is to apply DDRH theory to
asymmetric matter and nuclei far off stability. Since the vertex
functionals used in the former applications were taken from DB
calculations in symmetric matter, information on the density
dependence of isovector vertices was not available. From the DB
results for asymmetric matter, obtained in 
Ref.~\cite{deJong:98a,deJong:98b} with the Groningen potential
\cite{Malfliet:88}, a new set of coupling constants has been
derived, now including density dependent isoscalar ($\sigma$,
$\omega$) and isovector ($\delta$, $\rho$) vertices.
Contributions from the scalar-isoscalar $\delta$ meson are of
special interest at extreme neutron-to-proton ratios.

From inifinite matter calculations it was found that the momentum
dependence of self-energies has to be taken into account in order
to reproduce the underlying DBHF equation of state by DDRH
calculations. In a strict sense this means to go beyond the
static Hartree limit. A closer inspection, however, shows that
for a mean-field description it is sufficient to account for the
momentum dependence on an average level. Rather than using the
self-energies at the Fermi-surface \cite{Brockmann:90,deJong:98a}
a more appropriate method is to extract coupling constants from
self-energies averaged over the Fermi-sphere. This still leads to
static vertices but incorporating momentum dependent corrections.

The paper is arranged as follows. In Sec.~\ref{sec:DensDepNucMat}
the DDRH approach and the mean-field reduction are reviewed. In
Sec.~\ref{sec:NucMat} the approach to extract density dependent
coupling constants from nuclear matter DB self-energies including
isovector contributions and the momentum corrections is
presented. The global properties of the newly determined coupling
constants are investigated in applications to infinite matter. In
Sec.~\ref{sec:RelHarFinNuc} we report on DDRH calculations for
the isotopic chains of Ni and Sn nuclides, both including stable
and exotic nuclei. These two isotopic chains are of special
interest because several magic neutron numbers and the
corresponding shell closures are covered on top of the magic Z=28
and Z=50 proton numbers, respectively. Results for the Groningen
and the Bonn A NN potentials are compared. Contribution from the
scalar-isoscalar $\delta$ meson and the influence of the momentum
correction are examined in detail. The paper closes in
Sec.~\ref{sec:Summary} with a summary and conclusions.

\section{Density dependent hadron field theory for asymmetric nuclear
matter}
\label{sec:DensDepNucMat}
\subsection{The model Lagrangian and the equations of motion}
\label{ssec:Lagr}
The density dependent relativistic hadron field (DDRH) theory has
been presented and thoroughly discussed in \cite{Lenske:95,Fuchs:95,Keil:00}.
In this work we restrict ourselves to a short review of the model
and to a discussion of the extensions to Ref.~\cite{Fuchs:95}.

The model Lagrangian includes the baryons represented as Dirac spinors $\Psi =
\left(\psi_p,\psi_n\right)^T$, the isocalar mesons $\sigma$ and $\omega$, the
isovector $\rho$ meson and the photon $\gamma$. In addition to former models we
also include the scalar isovector meson $\delta$ which is important in
asymmetric systems and naturally has to be taken into account when extracting
the coupling functionals from asymmetric nuclear matter DBHF calculations as
will be explained in detail later. The Lagrangian is
\beqa
\Lcal &=& \Lcal_{B} + \Lcal_{M} + \Lcal_{int} \nonumber \\
\Lcal_{B} &=& \Psib \left[ i\gamma_\mu\p^\mu - M \right] \Psi \\
\Lcal_{M} &=&\frac{1}{2} \sum_{i=\sigma,\delta}
    \left(\p_\mu\Phi_i\p^\mu\Phi_i - m_i^2\Phi_i^2\right) - \nonumber
    \\
& & \frac{1}{2} \sum_{\kappa=\omega,\rho,\gamma}
    \left( \frac{1}{2} F^{(\kappa)}_{\mu\nu} F^{(\kappa)\mu\nu}
    - m_\kappa^2 A^{(\kappa)}_\mu A^{(\kappa)\mu}
    \right) \label{eq:Lagrangian} \\
\Lcal_{int} &=&
   \Psib\hat{\Gamma}_{\sigma}(\Psib,\Psi)\Psi\Phi_{\sigma} -
   \Psib\hat{\Gamma}_{\omega}(\Psib,\Psi)\gamma_{\mu}\Psi A^{(\omega)
   \mu} + \nonumber \\
&& \Psib\hat{\Gamma}_{\delta}(\Psib,\Psi)\bm{\tau}\Psi\bm{\Phi}
_{\delta} -
   \Psib\hat{\Gamma}_{\rho}  (\Psib,\Psi)\gamma_{\mu}\bm{\tau}
   \Psi\bm{A}^{(\rho)\mu} - \nonumber \\
&& e\Psib\hat{Q}\gamma_{\mu}\Psi A^{(\gamma)\mu}.
\eeqa
Here, $\Lcal_B$ and $\Lcal_M$ are the free baryonic and the free
mesonic Lagrangians, respectively, and interactions are described by
$\Lcal_{int}$, where
\beq F^{(\kappa)}_{\mu\nu} = \p_\mu A_\nu^{(\kappa)} - \p_\nu
A_\mu^{(\kappa)}
\label{eq:Fmunu}
\eeq
is the field strength tensor of either the vector mesons ($\kappa=
\omega,\rho$) or the photon ($\kappa=\gamma$) and $\hat{Q}$ is the
electric charge operator.

The main difference to standard QHD models \cite{Walecka:74,Serot:86} is that
the meson-baryon vertices $\hat\Gamma_\alpha$ ($\alpha=\sigma,\omega,\delta,
\rho$) are not constant numbers but depend on the baryon field operators $\Psi$.
Relativistic covariance requires that the vertices are functions
$\hat\Gamma_{\alpha}(\hat\rho)$ of Lorenz-scalar bilinear forms $\hat\rho(\Psib,
\Psi)$ of the field operators. Two obvious choices are the scalar density
dependence (SDD) with $\hat\rho=\Psib\Psi$ and the vector density dependence
(VDD) where $\hat\rho^2=\hat j_{\mu}\hat j^{\mu}$ depends on the square of the
baryon vector current $\hat j_{\mu}=\Psib\gamma_{\mu}\Psi$. In this work we only
present results for the VDD description since it leads to better results for
finite nuclei \cite{Fuchs:95} and gives a more natural connection to the
parameterization of the DB vertices. This will be discussed in detail in the
next section.

As pointed out in \cite{Fuchs:95} the most important difference to RMF
\cite{Gambhir:90} or convential DD \cite{Brockmann:92} theories is the
contribution from the rearrangement self-energies to the DDRH baryon
field equations. This is evident since the variational derivative of
$\Lcal_{int}$ with respect to $\Psib$ will also act on the vertices.
\beq
\label{eq:Variation}
\frac{\delta\Lcal_{int}}{\delta\Psib} = \frac{\p\Lcal_{int}}{\p\Psib}
+ \frac{\p\Lcal_{int}}{\p\hat\rho} \frac{\delta\hat\rho}{\delta\Psib}
\eeq
The second term on the right hand side of the equation is the
rearrangement contribution to the self-energy. Rearrangement accounts
physically for static polarization effects in the nuclear medium,
cancelling certain classes of particle-hole diagrams \cite{Negele:82}.
The usual self-energies are defined as
\beqa
\hat\se^{s(0)} & = &   \hat\Gamma_{\sigma}(\hat\rho)\Phi_{\sigma} +
                       \hat\Gamma_{\delta}(\hat\rho)\bm{\tau}\bm{\Phi}
                       _{\delta} \\
\hat\se^{\mu(0)} & = & \hat\Gamma_{\omega}(\hat\rho)A^{(\omega)\mu} +
                       \hat\Gamma_{\rho}  (\hat\rho)\bm{\tau}\bm{A}
                       ^{(\rho)\mu} +
                       e\hat Q A^{(\gamma)\mu}.
\eeqa
while the vector rearrangement self-energies are obtained from
Eq.~(\ref{eq:Variation}) as
\beqa
\hat\se^{\mu(r)} & = & \left(
   \frac{\p\hat\Gamma{_\omega}}{\p\hat\rho}A^{(\omega)\nu}\hat j_{\nu}
   +
   \frac{\p\hat\Gamma{_\rho}}{\p\hat\rho}\bm{\tau}\bm{A}^{(\rho)\nu}
   \hat j_{\nu}
\right. \nonumber \\ & - & \left.
   \frac{\p\hat\Gamma{_\sigma}}{\p\hat\rho}\Phi_{\sigma}\Psib\Psi -
   \frac{\p\hat\Gamma{_\delta}}{\p\hat\rho}\bm{\tau}\bm{\Phi}_{\delta}
   \Psib\Psi
                   \right) \hat u^{\mu}.
\eeqa
Here, $\hat u^{\mu}$ is a four velocity with $\hat u^{\mu}\hat
u_{\mu}=1$. Defining
\beq
\hat\se^s = \hat\se^{s(0)}, \quad \hat\se^{\mu} = \hat\se^{\mu(0)} +
\hat\se^{\mu(r)}
\eeq
the structure of the baryon field equations takes on the standard form
\beq
\left[\gamma_\mu \left( i\p^\mu - \hat\se^\mu \right) - \left( M -
\hat\se^s \right) \right] \Psi = 0,
\eeq
however, the underlying dynamics is changed by the rearrangement
contributions. In addition the effective baryon mass $M^* = M -
\hat\se^s$ differs for protons and neutrons due to the inclusion of
the scalar isovector $\delta$ meson. This is an additional property of
the model that was not present in the previous formulation
\cite{Lenske:95,Fuchs:95} where only the $\rho$ meson in the isovector
part of the interaction was considered. We also find that the density
dependence of the $\rho$ and $\delta$ mesons gives an additional
contribution to the vector rearrangement self-energies.

\subsection{Mean-field reduction}
\label{ssec:MFReduction}
The field equations are solved in the Hartree mean-field
approximation. In the Hartree approach the highly complex form of the
vertex functionals and its derivatives can be treated in a simple way
using Wick's theorem \cite{Wick:50}. Calculating the expectation value
with respect to the ground state $|0\rangle$ the vertices reduce to
\beq
\langle\hat\Gamma_{\alpha}(\hat\rho)\rangle = \Gamma_{\alpha}
(\langle\hat\rho\rangle)
= \Gamma_{\alpha}(\rho)
\eeq
and
\beq
\left<\frac{\p\hat\Gamma_{\alpha}(\hat\rho)}{\p\hat\rho}\right> =
\frac{\p\Gamma_{\alpha}(\rho)}{\p\rho}.
\eeq
where in the VDD case $\langle\hat\rho\rangle=\rho$ is just the baryon
ground state density.

Meson fields are treated as static classical fields, time reversal
symmetry is assumed, therefore only the zero component of the vector
fields contributes. The meson field equations reduce to
\beqa
(-\bm{\nabla}^2 + m_{\sigma}^2)\Phi_{\sigma}  &=&  \Gamma_{\sigma}
(\rho)\rho^s
\label{eq:MesonFields}\\
(-\bm{\nabla}^2 + m_{\omega}^2)A^{(\omega)}_0 &=&  \Gamma_{\omega}
(\rho)\rho \\
(-\bm{\nabla}^2 + m_{\delta}^2)\Phi_{\delta}  &=&  \Gamma_{\delta}
(\rho)\rho^s_3 \\
(-\bm{\nabla}^2 + m_{\rho}^2)A^{(\rho)  }_0   &=&  \Gamma_{\rho}(\rho)
\rho_3
\label{eq:MesonFielde}\\
-\bm{\nabla}^2 A^{(\gamma)}_0 &=& -e\rho_{p}
\label{eq:MesonFieldg}
\eeqa
where the densities are the following ground state expectation values
\beqa
\rho^s   = \langle\Psib\Psi\rangle &=& \rho_n^s + \rho_p^s
\label{eq:Densitiess}\\
\rho     = \langle\Psib\gamma_{0}\Psi\rangle &=& \rho_n + \rho_p \\
\rho_3^s = \langle\Psib\tau_{3}\Psi\rangle &=& \rho_n^s - \rho_p^s \\
\rho_3   = \langle\Psib\gamma_{0}\tau_{3}\Psi\rangle &=& \rho_n -
\rho_p
\label{eq:Densitiese}
\eeqa
and the indices $n$ and $p$ stand for neutrons and protons,
respectively. We will use the index $b=n,p$ to distinguisch between
different nucleons.

The Dirac equation, separated in isospin, is the only remaining
operator field equation
\beq
\label{eq:Dirac}
\left[\gamma_\mu \left( i\p^\mu - \se_b^\mu(\rho) \right) - \left( M -
\se_b^s(\rho) \right) \right]
\psi_b = 0
\eeq
and contains now the static density dependent self energies
\beqa
\se_b^{s(0)}(\rho) & = & \Gamma_{\sigma}(\rho)\Phi_{\sigma}
                     + \tau_b\Gamma_{\delta}(\rho)\Phi_{\delta} \\
\se_b^{0(0)}(\rho) & = & \Gamma_{\omega}(\rho)A_0^{(\omega)}
                     + \tau_b\Gamma_{\rho}  (\rho)A_0^{(\rho  )} +
                     e\frac{1-\tau_b}{2} A_0^{(\gamma)} \\
\se^{0(r)}(\rho) & = & \left(
   \frac{\p\Gamma{_\omega}}{\p\rho}A_0^{(\omega)}\rho +
   \frac{\p\Gamma{_\rho}}  {\p\rho}A_0^{(\rho)}\rho_3
\right. \nonumber \\ & - & \left.
   \frac{\p\Gamma{_\sigma}}{\p\rho}\Phi_{\sigma}\rho^s -
   \frac{\p\Gamma{_\delta}}{\p\rho}\Phi_{\delta}\rho^s_3
   \right).
\eeqa

The self-energies differ for protons and neutrons ($\tau_n=+1$,
$\tau_p=-1$) while the rearrangement self-energies are
independent of the isospin.

\section{Relativistic Hartree description of infinite nuclear matter}
\label{sec:NucMat}
\subsection{Properties of infinite nuclear matter}
\label{subsec:NucMatProp}

In nuclear matter the field equations further simplify assuming translational
invariance and neglecting the electromagnetic field. Solutions of the stationary
Dirac equation
\beq
\left[\gamma_{\mu}k_b^{*\mu}-m_b^*\right]u_b^*(k)=0
\eeq
are the usual plane wave Dirac spinors \cite{Bjorken:65}
\beq
u^*_b(k) = \sqrt{\frac{E^*_b+m^*_b}{2m^*_b}}
\left( \begin{array}{c} 1 \\ \frac{\bm{\sigma} \bm{k}^*_b}
{E^*_b+m^*_b}
       \end{array} \right) \chi_{b}
\eeq
where $\chi_b$ is a two-component Pauli spinor and the index $b$ distinguishes
between neutrons and protons. The effective mass $m^*_b=M-\se_b^s$ differs for
neutrons and protons due to the inclusion of the $\delta$ meson in the scalar
self-energy. The kinetic 4-momenta $k_b^{*\mu} = k_b^{\mu} - \se_b^{\mu}$ and
the energy $E_b^*$ of the particle are related by the in-medium on-shell
condition ${k_b^{*}}^2={m_b^{*}}^2$ leading to $E^*_b={(k_b^{*0})}^2 =
\sqrt{{\bm{k}^*_b}^2 + {m^*_b}^2}$. Integrating over all states $k\leq k_{F_b}$
inside the Fermi sphere and introducing $E_{F_b}=\sqrt{k_{F_b}^2+{m_b^*}^2}$ the
scalar and vector densities in infinite nuclear matter are found as
\beqa 
\rho_b & = &
\frac{2}{(2\pi)^3}\int_{|k|<k_{F_b}}d^3 k = \frac{k_{F_b}^3}{3\pi^2}
\label{eq:RhoB}\\
\rho_b^s & = & \frac{2}{(2\pi)^3}\int_{|k|<k_{F_b}}d^3 k \frac{m_b^*}
{E_b^*} \nonumber \\
         & = & \frac{m_b^*}{2\pi^2}\left[k_{F_b} E_{F_b}
           + {m_b^*}^2\ln\frac{k_{F_b}+E_{F_b}}{m_b^*}\right].
\label{eq:RhoS}
\eeqa

Calculation of the energy density and the pressure from the
energy-momentum tensor
\beqa
T^{\mu\nu} = \sum_i \frac{\p\Lcal}{\p(\p_{\mu}\phi_i)}\p^{\nu}\phi_i -
g^{\mu\nu}\Lcal
\\ \nonumber \phi_i = \Psib,\Psi,\Phi_{\sigma},A^{(\omega)}_{\mu},
\Phi_{\delta},A^{(\rho)}_{\mu}
\eeqa
is straightforward and the results are obtained in closed form
\beqa
\epsilon = \langle T^{00} \rangle & = & \sum_{b=n,p}
    \frac{1}{4} \left[ 3E_{F_b}\rho_b + m_b^*\rho_b^s \right]
    \nonumber \\ & + &
    \frac{1}{2} \left[ m_{\sigma}^2\Phi_{\sigma}^2
                     + m_{\delta}^2\Phi_{\delta}^2
                     + m_{\omega}^2{A^{(\omega)}_0}^2
                     + m_{\rho}^2{A^{(\rho)}_0}^2
                     \right] \nonumber \\
    & = &
    \sum_{b=n,p} \frac{1}{4} \left[ 3E_{F_b}\rho_b + m_b^*\rho_b^s
    \right] \nonumber \\ & + &
    \sum_{b=n,p} \frac{1}{2} \left[ \rho_b\se_b^{0(0)} +
    \rho_b^s\se_b^{s(0)} \right]
\label{eq:eNucMat} \\
p = \frac{1}{3}\sum_{i=1}^3\langle T^{ii} \rangle & = &
    \sum_{b=n,p} \frac{1}{4} \left[ E_{F_b}\rho_b - m_b^*\rho_b^s
    \right] +
    \sum_{b=n,p} \rho_b\se^{0(r)}  \nonumber \\ & - &
    \frac{1}{2}\left[m_{\sigma}^2\Phi_{\sigma}^2
                   + m_{\delta}^2\Phi_{\delta}^2
                   - m_{\omega}^2{A^{(\omega)}_0}^2
                   - m_{\rho}^2{A^{(\rho)}_0}^2
                   \right] \nonumber \\
    & = &
    \sum_{b=n,p} \frac{1}{4} \left[ E_{F_b}\rho_b - m_b^*\rho_b^s
    \right] +
    \rho\se^{0(r)}  \nonumber \\  & + &
    \sum_{b=n,p} \frac{1}{2} \left[\rho_b\se_b^{0(0)} -
    \rho_b^s\se_b^{s(0)}\right]
\label{eq:pNucMat}
\eeqa

From these relations it is seen that rearrangement does not affect the
energy density but contributes explicitly to the pressure $p$. It is
obvious from this that not taking into account rearrangement would
violate thermodynamical consistency because the mechanical pressure
$p$ obtained from the energy-momentum tensor must coincide with the
thermodynamical derivation
\beq
p_{\text{thermo}}=\rho^2\frac{\p}{\p\rho}\left(\frac{\epsilon}{\rho}
\right)=p.
\eeq

\subsection{Nucleon-meson vertices from Dirac-Brueckner theory and the
momentum correction}
\label{ssec:VertexMomentum}

Since the DDRH nucleon-meson vertices are deduced from microscopic
Dirac-Brueckner calculations the question arises about the best
\emph{ansatz} for the extraction of the DB results. This has been
discussed extensively in \cite{Keil:00}. For a given infinite nuclear
matter DB vertex $\Gamma_{\alpha}(\rho_{nm})$ the mapping to the field
theoretical formulation is defined by
\cite{Fuchs:95}
\beq \Gamma_{\alpha}(\hat\rho) = \int_0^{\infty}
\Gamma_{\alpha}(\rho_{nm})
       \delta(\rho_{nm}^2-\hat\rho^2)2\rho_{nm}d\rho_{nm}
\eeq
and directly allows us to apply the DB results to our model. Still, we have not
defined how to extract the DB vertices $\Gamma_{\alpha}(\rho_{nm})$ at a given
density $\rho_{nm}$ from the results obtained in Brueckner calculations. Results
of Brueckner calculations are the binding energy and the DB selfenergies
$\se^{\text{DB}}$. The latter are usually calculated by projecting the T matrix
onto a set of Lorentz invariant amplitudes 
\cite{Horowitz:87,Malfliet:88,McNeil:83}. 
They can then be related to coupling constants used in mean-field theory as
was examined in the local density approximation (LDA), e.g.~
\cite{deJong:98a,Brockmann:92,Haddad:93,Muether:90}. 
This is usually done on the level of the
infinite nuclear matter meson field equations by setting the mean-field self-
energies equal to $\se^{\text{MF}}=\Gamma_{\alpha}\phi_{\alpha}
\equiv\se^{\text{DB}}$. Plugging this into the meson field equations we find
$m_{\alpha}^2 \se^{\text{DB}}=\Gamma^2_{\alpha}\rho_{\alpha}$, with
$\rho_{\alpha}$ being the corresponding density to a meson field $\phi_{\alpha}$
as defined in Eqs.~(\ref{eq:Densitiess})-(\ref{eq:Densitiese}).

From Brueckner calculations in asymmetric nuclear matter scalar and vector self-
energies for protons and neutrons are given, allowing us to extract the
intrinsic density dependence of isocalar and isovector meson-nucleon vertices.
One finds \cite{deJong:98a}
\beqa \label{eq:maps}
\left( \frac{\Gamma_{\sigma}}{m_{\sigma}}\right)^2 & = & \frac{1}{2}
       \frac{\se^{s(DB)}_{n}(k_F) + \se^{s(DB)}_{p}(k_F)}{\rho_n^s +
         \rho_p^s} \\
\left( \frac{\Gamma_{\omega}}{m_{\omega}}\right)^2 & = & \frac{1}{2}
       \frac{\se^{0(DB)}_{n}(k_F) + \se^{0(DB)}_{p}(k_F)}{\rho_n +
         \rho_p} \\
\left( \frac{\Gamma_{\delta}}{m_{\delta}}\right)^2 & = & \frac{1}{2}
       \frac{\se^{s(DB)}_{n}(k_F) - \se^{s(DB)}_{p}(k_F)}{\rho_n^s -
         \rho_p^s} \\
\left( \frac{\Gamma_{\rho}}{m_{\rho}}\right)^2 & = & \frac{1}{2}
       \frac{\se^{0(DB)}_{n}(k_F) - \se^{0(DB)}_{p}(k_F)}{\rho_n -
       \rho_p}.
\label{eq:mape} \eeqa

From this follows that in general the vertices are functions of the Fermi
momentum and the scalar and vector densities. Specific parameterizations will be
discussed in the next section. In mean-field theory, only the ratios
$\frac{\Gamma_{\alpha}} {m_{\alpha}}$ determine the properties of the EoS. The
same still holds true for DDRH theory as long as the ratios
$\frac{\Gamma_{\alpha}}{m_{\alpha}}(k_F)$ are the same functions of $k_F$
\cite{Typel:99}. Comparing this approach to nonlinear RMF models 
\cite{Sugahara:94,Boguta:77,Bodmer:91}, 
one finds that the nonlinear $\sigma$ or $\omega$ terms
can also be interpreted as density dependent $\sigma$ or $\omega$ masses or vice
versa. However, for finite nuclei this is no longer correct since the
rearrangement dynamics alters the local single particle properties during the
self-consistent calculation. As a consequence mass and coupling strength
influence the system independently. In this paper we only consider constant
meson masses and put the medium dependence completely into the coupling
constants.

Self-energies of Brueckner calculations are in general momentum dependent. But
the usual approach is to neglect the momentum dependence and take the value at
the Fermi surface \cite{deJong:98a,deJong:98b} or to neglect it already \emph{a
priori} \cite{Brockmann:90}. Since the mapping is done on the Hartree level,
exchange contributions are implicitly parameterized into the direct terms. In
order to quantify the error from neglecting the momentum dependence, we expand
the full DB self-energies around the Fermi momentum.
\beqa \label{eq:kExpand}
\se^{\text{DB}}(k,k_F) & = & \se^{\text{DB}}(k_F,k_F) \nonumber \\
                & + & (k^2-k_F^2) \frac{\p \se^{\text{DB}}(k,k_F)}{\p
                k^2}|_{k=k_F}
                + O(k^4) \\
                & \equiv & \se^{\text{DB}}(k_F) + (k^2-k_F^2)
                \se'(k_F)
\eeqa

It is common practice to identify the first term with the Hartree
self-energy \cite{Brockmann:90}. A measure of the momentum dependence
around the Fermi surface is provided by the second term. A quadratic
dependence on the momentum has been chosen as supported by Brueckner
calculations \cite{deJong:98a}. However, using only $\se^{\text{DB}}
(k_F)$ for the determination of the vertices will, in general, not
reproduce the DB EoS. Up to now a satisfactory solution to this known
problem \cite{Haddad:93,Cescato:98} was not yet found. To tackle this
problem we introduce as an additional constraint that the self-
energies have to be chosen such that $\epsilon^{\text{DB}} \equiv
\epsilon^{\text{DDRH}}$.

As can be seen from Eq.~(\ref{eq:eNucMat}), the mean-field
contribution of the vector self-energy to the potential energy of
symmetric nuclear matter is given by $\epsilon_{pot}^0(k_F) =
\rho\se^{0(0)}(k_F)$. Averaging the same contribution from the DB
self-energies $\se^{\text{DB}}(k,k_F)$ over the Fermi sphere and
requiring it to equal the mean-field potential energy we find the
condition
\beqa
\rho(k_F) \se^{0(0)}(k_F) & = & \frac{4}{(2\pi)^3} \int_{|k|\leq k_F}
d^3k
\se^{\text{DB}}(k,k_F) \nonumber \\
                          & = & \rho(k_F) \se^{\text{DB}}(k_F) -
                          \se'(k_F)\frac{2}{15}k_F^5 \nonumber \\
                          & = & \rho(k_F) \se^{\text{DB}}(k_F)
                          \left[ 1 - \frac{2}{3}k_F^2 \frac{\se'(k_F)}
                          {\se^{\text{DB}}(k_F)} \right]
\eeqa

Obviously, the term in brackets is the correction that has to be taken into
account in order to reproduce the EoS. It should also be included in the
extracted self-energies. In principle, $\se'(k_F)$ is known from 
Eq.~(\ref{eq:kExpand}) 
but usually not extracted from DBHF calculations. Therefore, we use
the approach to calculate $\se'(k_F)$ numerically by adjusting the DDRH binding
energy to the DB EoS. This can also be interpreted as modifying the vertices
\beq \Gamma^2(k_F) \rightarrow \tilde{\Gamma}^2(k_F)
              \equiv \Gamma^2(k_F)\left[ 1 - \frac{2}{3}k_F^2
              \frac{\se'(k_F)}{\se^{\text{DB}}(k_F)} \right].
\eeq

As a first approximation we assume the ratio $\se'(k_F) /
\se^{\text{DB}}(k_F)$ to depend weakly on $k_F$ which motivates the
introduction of momentum corrected nucleon-meson vertices
\beq
\label{eq:modCouple}
\tilde{\Gamma}_{\alpha}(k_F) =  \Gamma_{\alpha}(k_F) \sqrt{ 1 +
\zeta_{\alpha} k_F^2 }
\eeq
with $\zeta_{\alpha}$ being constants determined by adjusting to the
DBHF EoS. The rearrangement terms are modified as follows
\beqa
\frac{\p \tilde{\Gamma}_{\alpha}(k_F)}{\p \rho}
  &=& \sqrt{1 + \zeta_{\alpha} k_F^2}  \frac{\p \Gamma_{\alpha}(k_F)}
  {\p \rho}
  \nonumber \\ &+& \frac{k_F}{3\rho} \frac{k_F \zeta_{\alpha}}{\sqrt{1
  + \zeta_{\alpha} k_F^2}}
  \Gamma_{\alpha}(k_F)
\eeqa
A more general \emph{ansatz} would be to let $\zeta_{\alpha}(k_F)$
depend on the Fermi momentum.

The contribution from the scalar mesons can be treated accordingly, however one
has to note that the scalar self-energy is also contained in the effective mass
$m^*$. Therefore a change of $\se^{s(0)}$ also affects $\rho_s$ and couples back
to the modified self-energies. For this reason it is not possible to give a
closed form for the exact momentum correction. Still, the \emph{ansatz} from
equation (\ref{eq:modCouple}) can be used to modify the scalar coupling
constants but with the constants $\zeta_{\alpha}$ to be fixed numerically.

It should be noted that the modified self-energies do not represent the exact DB
self-energies since the mapping was not done on the single particle level but by
adjusting the bulk binding energy of the EoS. Also, in order to have momentum
dependence on the single particle level, exchange terms would have to be taken
into account explicitly. However, these are already implicitly included in the
Hartree vertices through a Fierz transformation \cite{Serot:86,Keil:00}.
Nevertheless, bulk properties of the DB calculations are retained without
adjusting every single self-energy separately since $\zeta_{\alpha}$ was chosen
to be a constant thus keeping the number of new parameters on a minimal level.
The quality of this approximation is measured directly by the agreement of the
two equations of state.

One can imagine several possibilities how to calculate the momentum correction.
One way is to adjust $\zeta_{\sigma}$ and $\zeta_{\omega}$ to the minimum of
isospin symmetric infinite nuclear matter. Another way would be to keep e.g. the
$\sigma$ vertex fixed and adjust $\zeta_{\omega}$ for each point of the EoS
which leads to a density dependent correction $\zeta_{\omega}(k_F)$. Apparently,
the procedure to determine the momentum correction $\zeta_{\alpha}$ is not
unique, as already pointed out in \cite{deJong:98a,deJong:98b}.

In the next section we are going to present a parameterization of
coupling constants derived
from DB calculations in asymmetric nuclear matter and discuss results
of the momentum
correction by assuming $\zeta_{\alpha}=$ const.

\subsection{Fit of the nucleon-nucleon vertices}
\label{ssec:VertexFit}

\begin{figure}
\centering\epsfig{file=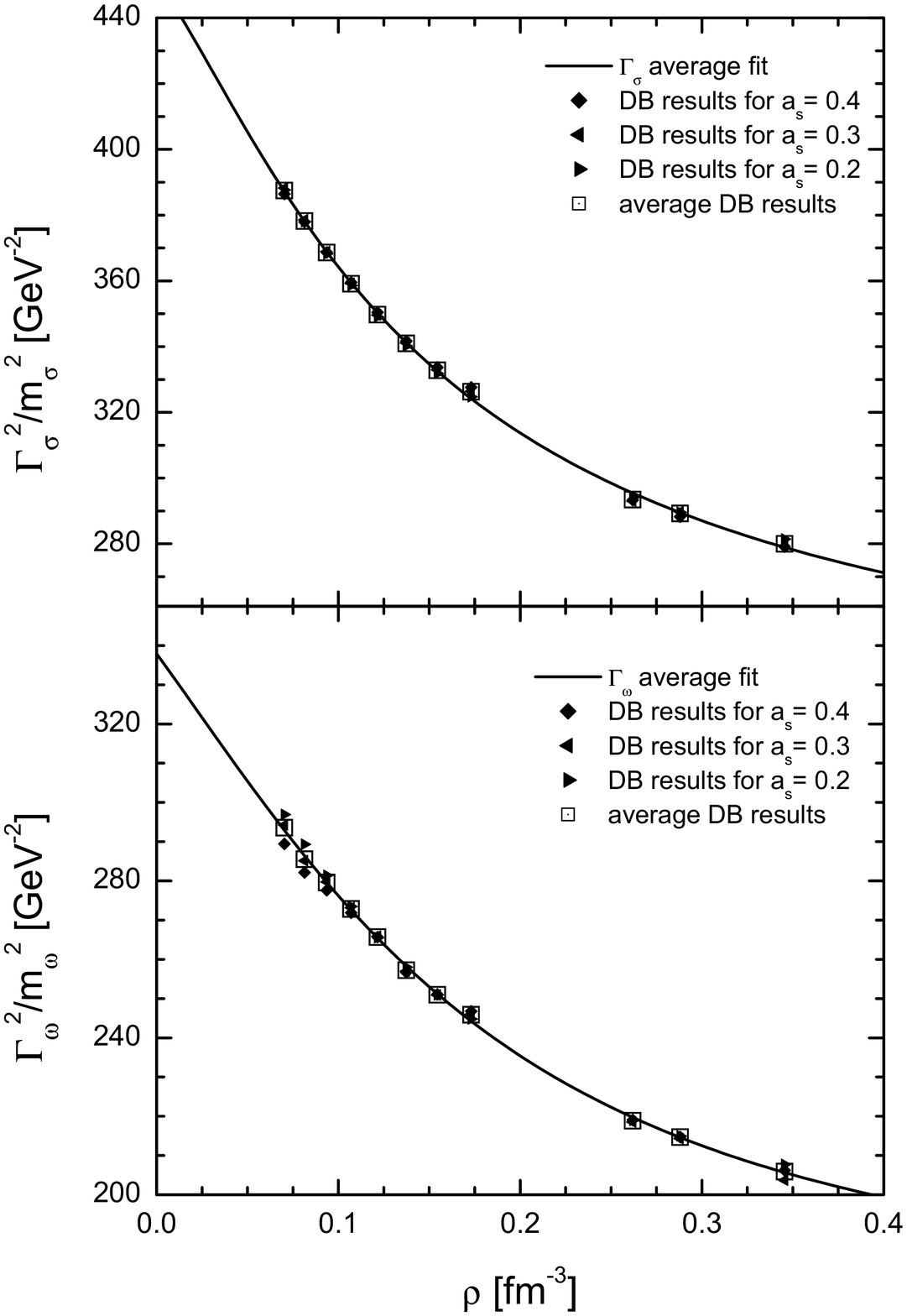,width=8.6cm}
\caption{Effective density dependence of the $\sigma$ (upper part) and
the $\omega$ (lower part) meson-nucleon vertices. Shown are results
extracted from DB selfenergies from the Groningen NN potential
\protect\cite{deJong:98a} calculated for the asymmetry ratios $a_s =
0.2, 0.3, 0.4$. The solid line is the asymmetry independent fit
through the average of the DB results (open squares).}
\label{fig:s-o-coupling}
\end{figure}

\begin{figure}
\centering\epsfig{file=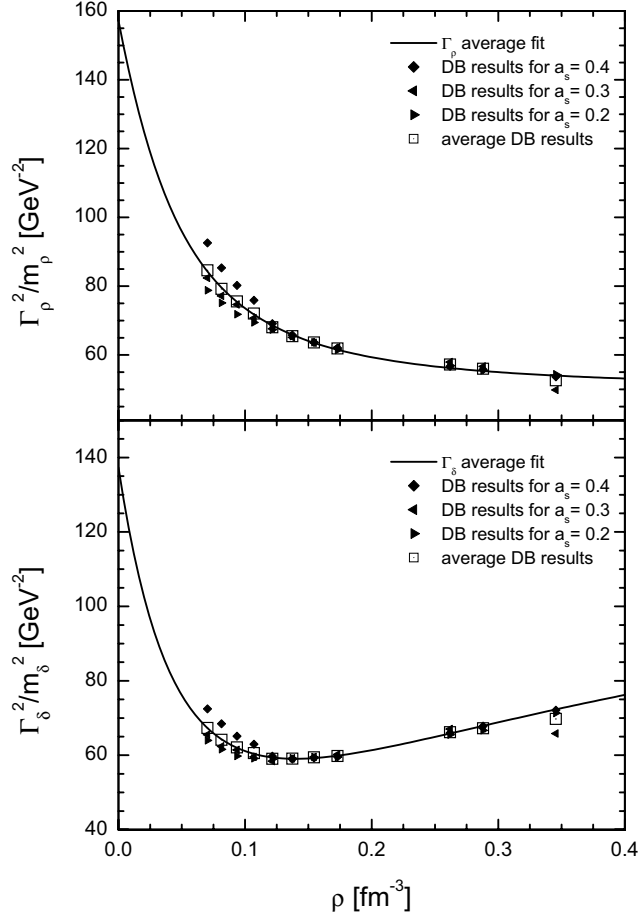,width=8.6cm}
\caption{Same as Fig.~\ref{fig:s-o-coupling} but for the $\rho$ (upper
part) and $\delta$ (lower part) meson-nucleon vertices.}
\label{fig:d-r-coupling}
\end{figure}

Several parameterizations of density dependent coupling constants exist. But
they either only include density dependence in the isoscalar channel
\cite{Haddad:93} due to the lack of asymmetric nuclear matter DB calculations or
they are purely phenomenological \cite{Typel:99}. Here we present a
parameterization of asymmetric nuclear matter results 
\cite{deJong:98a,deJong:98b}
derived from the Groningen potential \cite{TerHaar:87,Malfliet:88}. The
mapping of the DB self-energies is done as proposed in equations (\ref{eq:maps})
- (\ref{eq:mape}) leading to a density dependence in both the isoscalar
($\sigma$, $\omega$) and the isovector ($\delta$, $\rho$) channel.
Figure \ref{fig:s-o-coupling} and Fig.~\ref{fig:d-r-coupling} show the dependence
of the ratios $\frac{\Gamma_{\alpha}}{m_{\alpha}}(k_F)$ on the density
$\rho(k_F)$ for different asymmetry ratios $a_{s}=\rho_p / \rho$. In the
isoscalar channel the dependence on the asymmetry is negligible, in the
iscovector channel it is extremely weak, especially around the saturation
density of about $\rho_0 = 0.16 fm^{-3}$. We therefore choose the \emph{ansatz}
that the coupling constant only depend on the total vector density $\rho(k_F)$
and not on the proton and neutron densities separately. In order to take into
account a maximum of information from the DB calculations we fit the average of
the self-energies for the asymmetry ratios $a_s=0.2,0.3,0.4$ \cite{deJong:98a}.
In \cite{Haddad:93} a polynomial expansion of $\Gamma^2$ in $k_F$ around the
saturation density $k_{F_0}$ was chosen, leading to an excellent fit of the
self-energies around $\rho_0$. But due to the polynomial approach the behavior
at very low and very high densities is not perfectly stable. Intending
applications over wide density regions ranging from nuclear halos to neutron
star conditions in future investigations we choose a rational approximation as
proposed in Ref.~\cite{Typel:99}
\beq \label{eq:DDrat}
\Gamma_{\alpha}(\rho) = a_{\alpha}\left[
                        \frac{1+b_{\alpha}(\frac{\rho}{\rho_0}
                        +d_{\alpha})^2}
                             {1+c_{\alpha}(\frac{\rho}{\rho_0}
                             +e_{\alpha})^2}\right].
\eeq
A clear advantage of such a rational form is the well defined behavior
at low and high densities turning into a constant at very high
densities. The results for the fit for $\rho_0 = 0.16 fm^{-3}$ are
displayed in Fig.~\ref{fig:s-o-coupling} for the isoscalar channel and
in Fig.~\ref{fig:d-r-coupling} for the isovector channel of the
interaction. The parameters are shown in Table \ref{tab:coupling}. The
description of the DB results is very good, in the isoscalar channel
it is even sufficient to require $d_{\alpha}=e_{\alpha}$. In the
isovector channel the fit is more difficult, especially since the
$\delta$-meson has an ascending slope at high densities. This requires
the additional parameter $e_{\alpha}$ to describe the low and the high
density behavior equally well.

\subsection{Results for infinite nuclear matter}
\label{ssec:NucMatResults}

\begin{figure}
\centering\epsfig{file=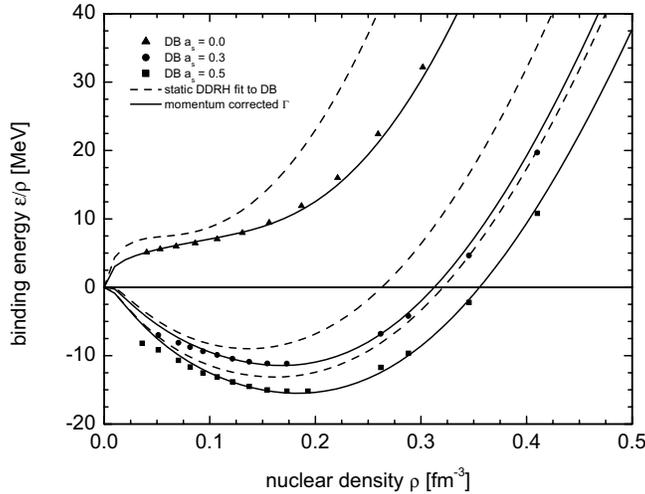,width=8.6cm}
\caption{Equation of state for different asymmetry ratios $a_s$
calculated from the Groningen NN potential. DB results from
\protect\cite{deJong:98a} are represented by symbols. The dashed line
shows the DDRH EoS for meson-nucleon vertices fitted to the DB self-
energies, the solid line denotes results with momentum correction.}
\label{fig:eos_groningen}
\end{figure}

To check the quality of the effective parameterization (Table
\ref{tab:coupling}) of the in-medium dependence of the vertices it is
instructive to look at the infinite nuclear matter EoS. The
calculation was done by solving the meson field equations 
(\ref{eq:MesonFields}-\ref{eq:MesonFielde}) 
with the densities from equations
(\ref{eq:RhoB}) and (\ref{eq:RhoS}). Results for symmetric nuclear
matter ($a_s=0.5$), pure neutron matter ($a_s=0.0$) and for nuclear
matter with an asymmetry ratio of $a_s=0.3$ are shown in
Fig.~\ref{fig:eos_groningen}. Displayed are also the DB binding
energies from \cite{deJong:98a} for the same NN potential. One sees
that the equation of state is clearly not reproduced even though the
fit describes the self-energies at $k_F$ very well. While the DB EoS
has a binding energy of $\epsilon/\rho_0=-15.6$ MeV and a saturation
density $\rho_0=0.182$ fm$^{-3}$, the standard choice leads to a
Hartree EoS which is about 2.5 MeV weaker bound ($\epsilon/\rho_0=-
13.13$ MeV) and the saturation density is shifted to lower densities
($\rho_0=0.161$ fm$^{-3}$). As discussed in 
Sec.~\ref{ssec:VertexMomentum} 
due to the approximations made when neglecting the
momentum dependence of the self-energies this could be expected.

\begin{figure}
\centering\epsfig{file=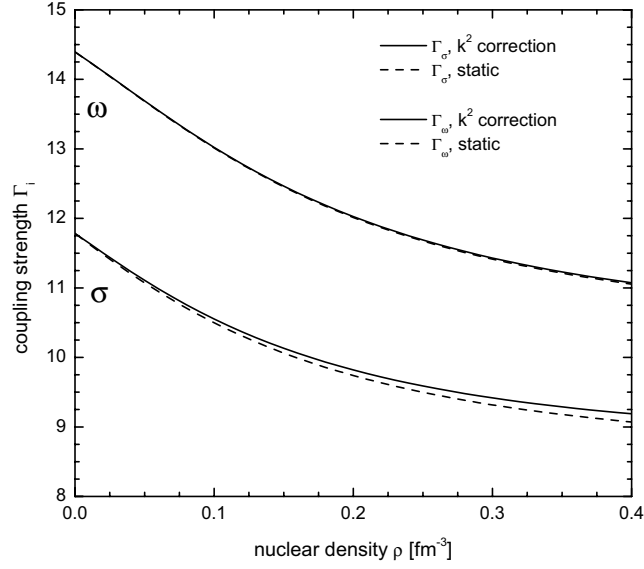,width=8.6cm}
\caption{Momentum corrected density dependence of the $\sigma$ and
$\omega$ meson-nucleon vertices. Results are shown for
$\zeta_{\sigma}=0.00804$ fm$^{-2}$ and $\zeta_{\omega}=0.00103$ 
fm$^{-2}$ (solid line) and compared to the fits from 
Fig.~\ref{fig:s-o-coupling} (dashed line).}
\label{fig:s-o-momentum-corr}
\end{figure}

We apply our momentum correction scheme to the coupling constants in a
two step process. First we restrict ourselves to symmetric nuclear
matter and try to reproduce the DB EoS by adjusting $\Gamma_{\sigma}$
and $\Gamma_{\omega}$. This is done in the constant momentum
correction scheme by choosing $\zeta_{\sigma}$ and $\zeta_{\omega}$ in
such a way that the saturation point of DB calculations is reproduced.
We find the very small corrections $\zeta_{\sigma}=0.00804$ fm$^{-2}$
and $\zeta_{\omega}=0.00103$ fm$^{-2}$ and are able to reproduce the
EoS very accurately as can be seen in Fig.~\ref{fig:eos_groningen}. It
is important to note that even though we only adjusted one point we
are able to reproduce the binding energies at low as well as at high
densities. This justifies our assumption of a $k_F^2$ dependence for
the correction of the coupling constants. 
Figure \ref{fig:s-o-momentum-corr} 
compares the momentum corrected couplings to the original ones.
The correction increases with higher densities (or momenta) as
expected from the functional form of the coupling constants but
remains small for the $\sigma$ meson and nearly negligible for the
$\omega$ meson. Nevertheless these small corrections suffice to gain
2.5 MeV binding energy at saturation density. We conclude that the EoS
reacts extremely sensitive to small changes in the coupling constants,
therefore great care has to be taken when fitting the self-energies.
In addition the same is appropriate for DB calculations. One has to be
very careful when extracting the self-energies and needs a consistent
scheme for the projection onto the Lorentz invariants.

\begin{figure}
\centering\epsfig{file=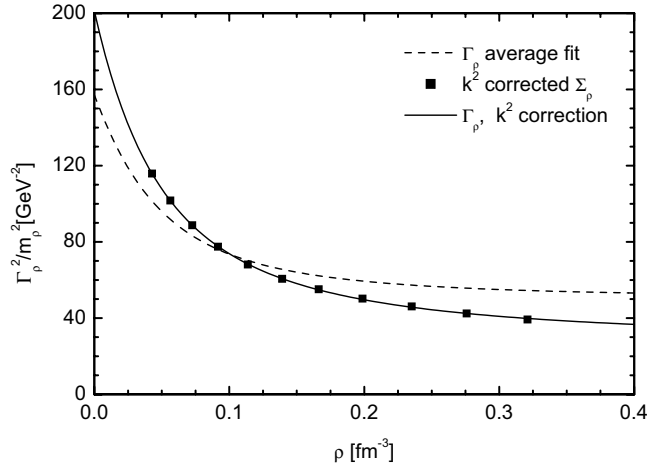,width=8.6cm}
\caption{Momentum corrected density dependence of the $\rho$ meson-
nucleon vertex. Results are shown for the fit (solid line) through the
corrected self-energies and compared to the fit from 
Fig.~\ref{fig:d-r-coupling} (dashed line).}
\label{fig:r-coupling-corr}
\end{figure}

After fixing $\Gamma_{\sigma}$ and $\Gamma_{\omega}$ the second step
is to adjust the couplings in the isovector channel. This is done by
keeping $\Gamma_{\delta}$ fixed and adjusting $\zeta_{\rho}$ for each
given DB binding energy to neutron matter. In this approach one
obtains a density dependent correction $\zeta_{\rho}(k_F)$. The
correction is incorporated in the DB self-energies and the $\rho$
meson-nucleon vertex is readjusted. The corrected self-energies and
the fit through them are shown in Fig.~\ref{fig:r-coupling-corr}, the
parameters are given in Table \ref{tab:RhoAdjusted}. From
Fig.~\ref{fig:eos_groningen} it is seen that the new fit reproduces
the EoS of neutron matter very well and this even at high densities
where the static fit to the DBHF self-energies leads to an interaction
being far too repulsive. This is important for applications to neutron
stars where the high density behavior plays a crucial role.
Figure \ref{fig:eos_groningen} also shows an excellent accordance of the
calculations with DB results at intermediate asymmetry ratios, e.g.
$a_s=0.3$, especially around the saturation density. This is a very
important result since the interaction was only adjusted to pure
neutron matter and justifies our assumption that the parameterization
of meson-nucleon vertices is asymmetry independent. We also confirmed
this for other values of $a_s$ where DB results were available from
\cite{deJong:98a}.

In Table \ref{tab:NucMatResults} nuclear matter properties for the
presented models are given. Saturation density and binding energy of
the momentum corrected DDRH calculation reproduce the DB data very
well, also the asymmetry-energy coefficient $a_4=26.1$ MeV, determined
by
\beq
a_4=\rho_0^2\frac{\p}{\p(\rho_{3})^2}\frac{\epsilon}{\rho}(\rho_0,
\rho_{3})|_{\rho_{3}=0},
\eeq
is in compliance with DB value of 25 MeV even though this value was
not taken into account for the adjustment of the isovector
interaction.

\section{Relativistic Hartree Description of finite nuclei}
\label{sec:RelHarFinNuc}
\subsection{Properties of finite nuclei}
\label{ssec:FiniteProp}

The density dependent interaction derived in the preceding section for nuclear
matter is now applied to finite nuclei in Hartree calculations. We solve the
full meson field equations (\ref{eq:MesonFields}-\ref{eq:MesonFieldg}) and the
Dirac equation (\ref{eq:Dirac}) in coordinate space. The Dirac equation is
solved for the upper and lower components of the eigenspinor $\psi_b$
simultaneously. The set of coupled equations is solved self-consistently under
the assumption of spherical symmetry. Pairing effects in the particle-particle
(p-p) channel have to be taken into account in open shell nuclei. Since we are
mainly interested in the mean-field particle-hole (p-h) channel and especially
in the isovector properties of the interaction, the BCS approximation was used.
This is a standard procedure in relativistic and non-relativistic mean-field
approaches. Following \cite{Brown:98} a constant pairing matrix element of $G=
2.15 \text{ MeV}/\sqrt{A}$ was assumed and the standard set of BCS equations
\cite{Preston:75} was solved independently for protons and neutrons,
respectively. In exotic nuclei close to the dripline pairing effects can be very
important if the Fermi energy is close to the continuum. Here, relativistic
Hartree-Bogoliubov (RHB) calculations, taking into account the coupling of bound
states to the continuum, have been performed for phenomenological interactions
\cite{Lalazissis:98,Meng:98,Carlson:99} leading to excellent agreement with
experimental results. We found in our calculations that pairing gives only minor
to negligible contributions compared to the effects from the microscopic
interaction in the p-h channel (less than 2\% to the total binding energy of the
Ni and Sn isotopes).

The center-of-mass correction which gives a significant contribution
to the binding energy of light nuclei is treated in the usual harmonic
oscillator approximation
\beq
E_{cm} = -\frac{3}{4}\hbar\omega \quad
\text{with} \quad \hbar\omega=41 A^{-1/3} \text{MeV}.
\eeq

Then, the total ground state energy that has to be compared with
experimental data is given by
\beq E_{g.s.} = E_{\text{MF}} + E_{pair} + E_{cm} \eeq where the
Hartree ground state energy
is obtained from the energy momentum tensor through spatial
integration of its $T^{00}$
component.
\beqa
E_{\text{MF}} & = & \sum_{i,\epsilon_i\leq\epsilon_F} v_i^2\epsilon_i
      - \int d^3r \rho(r)\se^{0(r)}(r) \nonumber \\ & &
      - \sum_{b=p,n} \frac{1}{2}\int d^3r
      \left[ \rho_b^s(r)\se_b^{s(0)}(r) - \rho_b(r)\se_b^{0(0)}(r)
      \right]
      \nonumber \\ & & + \frac{1}{2}\int d^3r\rho_p(r) e A^{(\gamma)}
      _0(r).
\eeqa
The $\epsilon_i$ are the Dirac eigenvalues of particles in positive energy
eigenstates and energies less or equal the Fermi energy $\epsilon_F$ and the
$v_i^2$ are the occupation probabilities obtained from BSC pairing.

In order to examine the effects of the density dependent isovector coupling
constants, it is instructive to look at the mean-field potentials. In leading
non- relativistic order the effective central potential $U_b^C$ is given by the
difference of the strongly attractive scalar and repulsive vector fields

\beq
U_b^C=\se_b^0 - \se_b^s
\eeq
while the strength of the spin-orbit potential
\beq
U_b^{SO}=\frac{1}{2M}\frac{-\p_{r}\left(\se_b^s+\se_b^0\right)}
                            {E+M-\left(\se_b^s+\se_b^0\right)}
\eeq
is determined by the sum of these fields.

Interestingly, the spin-orbit potential differs for protons and neutrons since
the self-energies depend on the isospin. We define the isoscalar and isovector
spin-orbit potentials as
\beqa
U_0^{SO} & = & \frac{1}{2}\left(U_n^{SO}+U_p^{SO}\right) \\
U_{\tau}^{SO} & = & \frac{1}{2}\left(U_n^{SO}-U_p^{SO}\right)
\eeqa
as a measure for isovector spin-orbit interactions. We expect to see
an enhancement of the isovector potential $U_{\tau}^{SO}$ due to the
inclusion of the $\delta$ meson. While, to a large extend, the
contributions of the $\rho$ and the $\delta$ mesons compensate each
other in the central potential $U_b^C$, producing an effective
isovector potential that is comparable in strength to the one obtained
in calculations that include only the $\rho$ meson, the isovector
self-energies of the $\rho$ and $\delta$ mesons add up and increase
the isovector spin-orbit potential $U_{\tau}^{SO}$. This will be
discussed in detail in the next section.

\subsection{Closed shell nuclei}
\label{ssec:ClosedShell}

As a first test of the momentum corrected density dependent
interaction we examine closed shell nuclei.
Nuclei considered in our calculations were the doubly magic nuclei
$^{16}$O, $^{40}$Ca, $^{100}$Sn, $^{132}$Sn, $^{208}$Pb, having major
shell gaps at 8, 20, 50 and 82 for both protons and neutrons, as well
as nuclei where one or both types of nucleons have only a minor
(semimagic) shell gap at 28 or 40 ($^{48}$Ca, $^{48}$Ni, $^{56}$Ni,
$^{68}$Ni, $^{90}$Zr). The nuclei have measured binding energies with
very small errors \cite{Audi:93} except for the proton rich nuclei.
The binding energy of $^{100}$Sn has been measured with a relatively
large error \cite{Chartier:96}. $^{48}$Ni, whose binding energy can be
extrapolated in terms of the mirror binding energy difference to
$^{48}$Ca \cite{Brown:98}, has recently been produced experimentally
\cite{Blank:00}. The set of nuclei includes stable as well as $\beta$
unstable nuclei covering the range from the proton dripline to very
neutron rich isotopes and allows us to investigate the interaction,
determined from asymmetric nuclear matter, extensively on finite
nuclei.

\subsubsection{Binding energies and charge radii}
\label{sssec:BindCharge}

\begin{figure}
\centering\epsfig{file=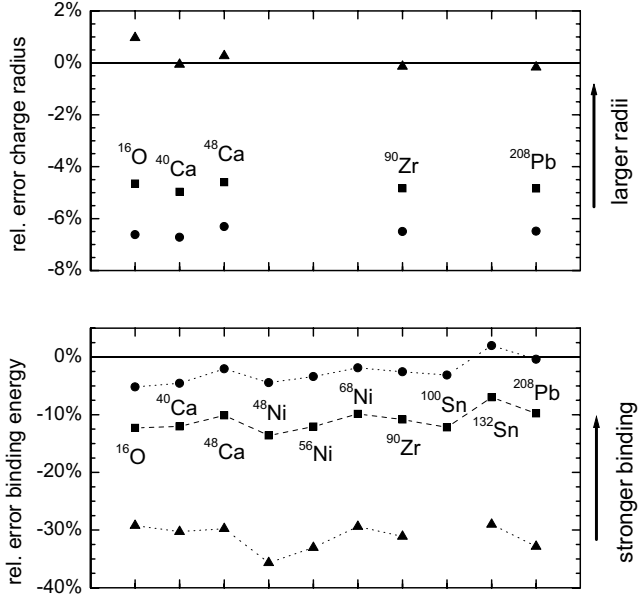,width=8.6cm}
\caption{Relative errors for charge radii $(\rho_c-
\rho_c^{\text{exp}})/\rho_c^{\text{exp}}$ (upper part) and binding
energies $(E_B-E_B^{\text{exp}})/E_B^{\text{exp}}$ (lower part)
obtained with the Groningen parameterization. Shown are results for
magic and semimagic nuclei. Results without momentum correction are
denoted by upper triangles, results for $\zeta_{\sigma} = 0.00804$
fm$^{-2}$ and $\zeta_{\omega} = 0.00103$ fm$^{-2}$ (adjusted to
nuclear matter) by squares and results for $\zeta_{\sigma} = 0.008$
fm$^{-2}$ and $\zeta_{\omega} = -0.002$ fm$^{-2}$ (adjusted to
finitite nuclei) by circles. The lines are drawn to guide the eye.}
\label{fig:groningen_magic}
\end{figure}

In Table \ref{tab:MagicResultsGro} we display results for closed shell nuclei
calculated with the DDRH parameterization of the Groningen NN potential derived
in Sec.~\ref{sec:NucMat}. Figure \ref{fig:groningen_magic} shows the relative
error for charge radii and binding energies compared to experimental results.
The parameterization derived directly from the self-energies without momentum
correction completely fails to describe the experimental results. While the
charge radii are described very well, all nuclei are strongly underbound by
about 30\%. The proton dripline of the tin isotopes is already reached before
$^{100}$Sn due to the weak binding of the protons. But this had to be expected
from the infinite nuclear matter results where the binding energy at saturation
is also too weak. Applying the momentum correction from 
Sec.~\ref{ssec:NucMatResults}, 
the results are greatly improved. While the binding energies are
still underestimated by about 10\%, compared to the static parameterization the
improvement is remarkable. On the other hand, the size of the charge radii
decreases, but with an error of about 4\% the general agreement with
experimental data is still satisfying.

Comparing our results with density dependent calculations for the Bonn A
interaction \cite{Fuchs:95,Ineichen:95,Cescato:98}, especially the description
of the binding energies is less satisfactory, but the reason for this lies in
the different NN potentials. While the parameterization of the Bonn A NN
potentials has at a nuclear matter saturation density of $\rho_0=0.162$ fm$^{-3}
$ an energy of about $-16.3$ Mev and for this reason tends to overbind finite
nuclei, the saturation point of the Groningen NN potential is shifted to higher
densities and the saturation energy is only about $-15.6$ MeV. These properties
are clearly translated to finite nuclei, leading for the Groningen interaction
to an underestimation of the binding energy, and a high nucleon density and deep
mean-field potential inside the nuclei. The high saturation density stronger
localizes the protons inside the nuclei leading to smaller charge radii. We
compared the theoretical charge density to experimental results and found an
overestimation of about 5\% to 10\% for all considered nuclei. Therefore, to
improve results for finite nuclei, Brueckner calculations and NN potentials have
to be improved first.

We were also interested in seeing how sensitive results for finite nuclei are to
the momentum correction and if it is possible to improve results by adjusting
the momentum correction factors $\zeta_{\alpha}$ to finite nuclei. No attempt
was made to optimize the parameters in order to obtain a perfect fit of data.
Isovector coupling constants remained unchanged. We found that a correction of
$\zeta_{\sigma}=0.008$ fm$^{-2}$ and $\zeta_{\omega}=-0.002$ fm$^{-2}$ provides
a reasonable description of both charge radii and binding energies. This
modification essentially corresponds to a weakening of the repulsion of the
vector self-energy compared to the original nuclear matter parameterization, but
other modifications lead to similar results. It should be noted that our
discussion of the momentum correction in infinite nuclear matter only partly
applies to the $\zeta_{\alpha}$ determined for finite nuclei and that the strict
relation to the factor $\se'/\se^{\text{DB}}$ is not retained. Obviously, we
implicitly take into account higher order effects and especially correct for the
deficiencies of the NN potential in reproducing the binding energies of finite
nuclei. This is demonstrated by the reversed sign of $\zeta_{\omega}$ compared
to the nuclear matter value.

Results are shown in Table \ref{tab:MagicResultsGro}. As can be seen from
Fig.~\ref{fig:groningen_magic}, charge radii are further decreased while the
agreement with the experimental binding energies is satisfying. In general, an
improvement in the binding energies leads to smaller values for the charge radii
and it is not possible to reproduce both observables simultaneously with the
same Groningen parameter set. An improvement of the results would be possible by
adjusting the $\sigma$ mass to better reproduce the charge radii but a
phenomenological fit to experimental results was not the goal of our
investigations.

\begin{figure}
\centering\epsfig{file=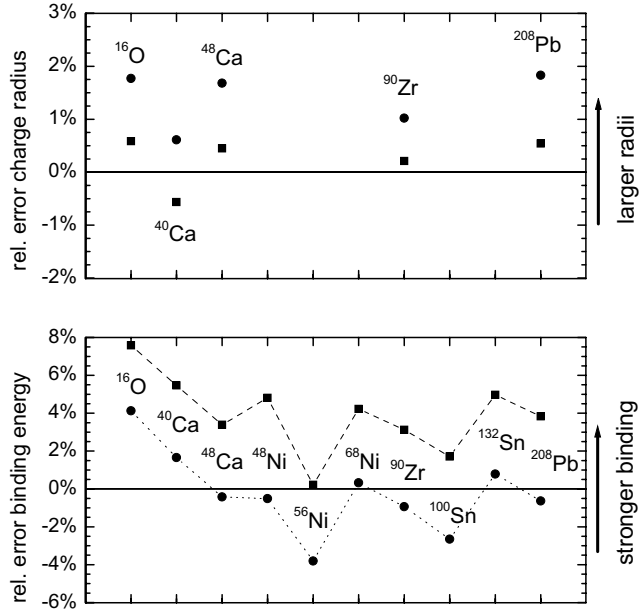,width=8.6cm}
\caption{Same as Fig.~\protect\ref{fig:groningen_magic} but for the
Bonn A parameterization from \protect\cite{Haddad:93,Fuchs:95}.
Results without momentum correction are denoted by squares and results
for $\zeta_{\sigma}=-0.003$ fm$^{-2}$ and $\zeta_{\omega}=-0.0015$
fm$^{-2}$ by circles. The lines are drawn to guide the eye.}
\label{fig:bonna_magic}
\end{figure}

We also investigated if the momentum correction could possibly improve the
description of finite nuclei for the Bonn nucleon potential. We used the
parameterization of the Bonn A $\sigma$ and $\omega$ meson-nucleon vertices from
\cite{Haddad:93} that was also applied to a few closed shell nuclei in
\cite{Fuchs:95,Ineichen:95}. The $\rho$ coupling strength is chosen as $g_{\rho}
^2/4\pi=5.19$ with a mass of $m_{\rho}=770$ MeV. Adjusting the momentum
correction factors to the values $\zeta_{\sigma}=-0.0030$ fm$^{-2}$ and
$\zeta_{\omega}=-0.0015$ fm$^{-2}$ leads to a good agreement with the
experimental binding energies (except for some $N= Z$ nuclei where p-n pairing
might be important) while preserving the good results for the charge radii.
Results are presented in Table \ref{tab:MagicResultsBoA} and 
Fig.~\ref{fig:bonna_magic}. 
One realizes the same behavior as for the Groningen
parameterization. Now, a decrease in the binding energies leads to an
overestimation of the charge radii.

The modification of the density dependent vertices is very small. Comparing the
original fit of Haddad and Weigel \cite{Haddad:93} to the momentum corrected
vertices we found that, in the density range important for finite nuclei, the
reproduction of the nuclear matter self-energies is of the same accuracy. This
leads to the conclusion that the momentum correction is clearly able to improve
results and confirms our statement from Sec.~\ref{sec:NucMat} that bulk
properties are very sensitive to small changes in the vertices.

We also recalculated the equation of state for the Bonn A potential with the
modified vertices. Comparing to the original parameterization we found the
agreement with the DB binding energy at saturation density to be slightly
improved but not perfect. At low densities both parameterizations fail to
reproduce the EoS. This seems to be a problem of the calculations of Brockmann
and Machleidt who assumed \emph{a priori} momentum-independent self-energies and
fitted them only to the positive-energy matrix elements of the scattering
matrix. We tried to adjust the coupling constants with our momentum correction
procedure but were unable to resolve the inconsistency between the DB 
self-energies and the DB binding energy, 
e.g. to reproduce the EoS for all densities.
This is not surprising considering the method of the DB calculations and the
fact that the momentum correction is not able to change the global properties of
the density dependence of the self-energies.

As can be seen from Figs. \ref{fig:groningen_magic} and \ref{fig:bonna_magic}
the momentum correction always leads to a shift of the relative errors that is
nearly identical across all considered nuclei. The reason is that the general
properties of the interaction were not altered, these are already determined by
the density dependence of the meson-nucleon vertices. Also the isovector part of
the interaction remained unchanged. We conclude that the momentum correction is
able to improve the extraction of the self-energies from Brueckner calculations
preserving the microscopic structure of the NN interaction.

\subsubsection{Spin-orbit splitting}
\label{sssec:SpinOrbit}

\begin{figure}
\centering\epsfig{file=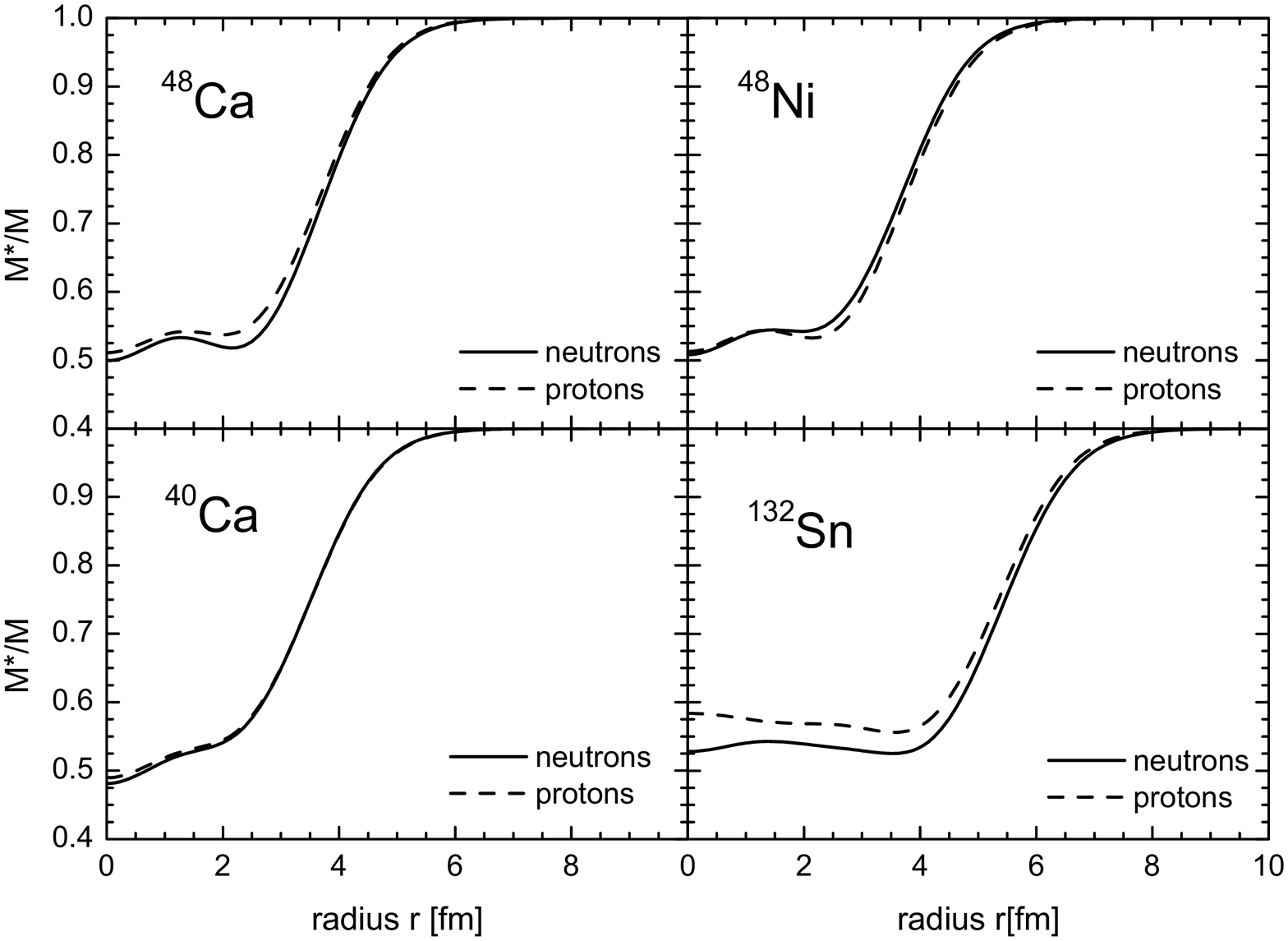,width=8.6cm}
\caption{Isospin dependence of the effective masses of closed-shell
nuclei from relativistic density dependent Hartree calculations using
the parameterization of the Groningen NN potential.}
\label{fig:eff_mass}
\end{figure}

As stated before, we expect the effects of including the $\delta$ meson to
manifest in the isovector spin-orbit potential and in isovector dependent
effective masses. In Fig.~\ref{fig:eff_mass} effective masses for a
representative sample of nuclei are displayed. While in the $N=Z$ nucleus $^{40}
$Ca the difference between proton and neutron masses is induced solely by
Coulomb effects and is negligible, in the neutron rich nucleus $^{132}$Sn the
proton and neutron effective masses differ by about 10\% at central density.
This enhancement is caused by the isovector contribution to the scalar self-
energy. The mirror nuclei $^{48}$Ca and $^{48}$Ni nicely demonstrate the effect
of the scalar isovector density by showing a reversed behavior in the effective
masses. We found that the momentum correction does not strongly effect the
effective masses, only slightly decreasing (increasing) them due to a deeper
(flatter) core potential.

\begin{figure}
\centering\epsfig{file=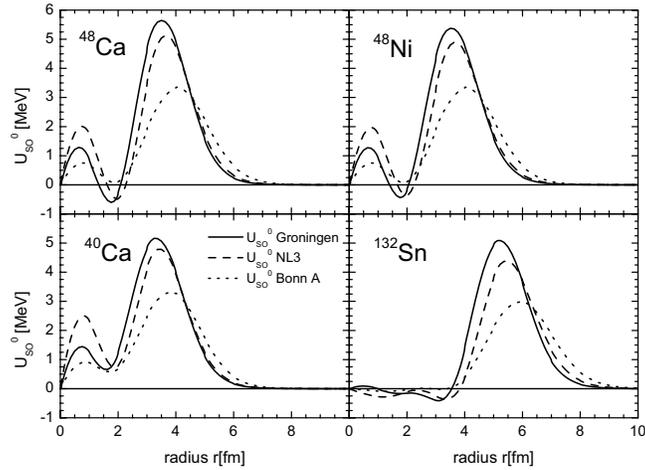,width=8.6cm}
\caption{Spin-orbit potentials of closed-shell nuclei from DDRH
calculations. Shown are the isoscalar potentials $U_0^{SO}$ for the
Groningen parameterization (solid lines) and the Bonn A
parameterization (dotted line). For comparison results calculated with
the phenomenological NL3 RMF parameter set are shown (dashed lines).}
\label{fig:spin-orbit}
\end{figure}

The effective masses calculated with the Groningen potential are quite small,
leading, together with the relatively large self-energies, to a large 
spin-orbit potential as can be seen from Fig.~\ref{fig:spin-orbit}. 
Analogous to the
effective masses no strong dependence on the momentum correction was found. For
comparison we also display the spin-orbit potential of the Bonn A and the
phenomenological NL3 parameter set \cite{Lalazissis:97}. The Bonn A spin-orbit
potential is too small with the peak structure at the nuclear surface shifted to
larger radii. We found the same behavior for the effective mass, being larger
and less localized in the center of the nucleus. The properties of the Groningen
and NL3 spin-orbit potentials are relatively similar, but the spin-orbit
splitting of the Groningen potential is too large (about a factor two larger
than the one obtained with Bonn A) as can be seen from Table 
\ref{tab:spin-orbit}. 
Compared to the NL3 parameterization that describes the experimental
values well, $\Delta_{LS}$ is about 25\% too strong, while for Bonn A it is
about 30\% too weak.

\begin{figure}
\centering\epsfig{file=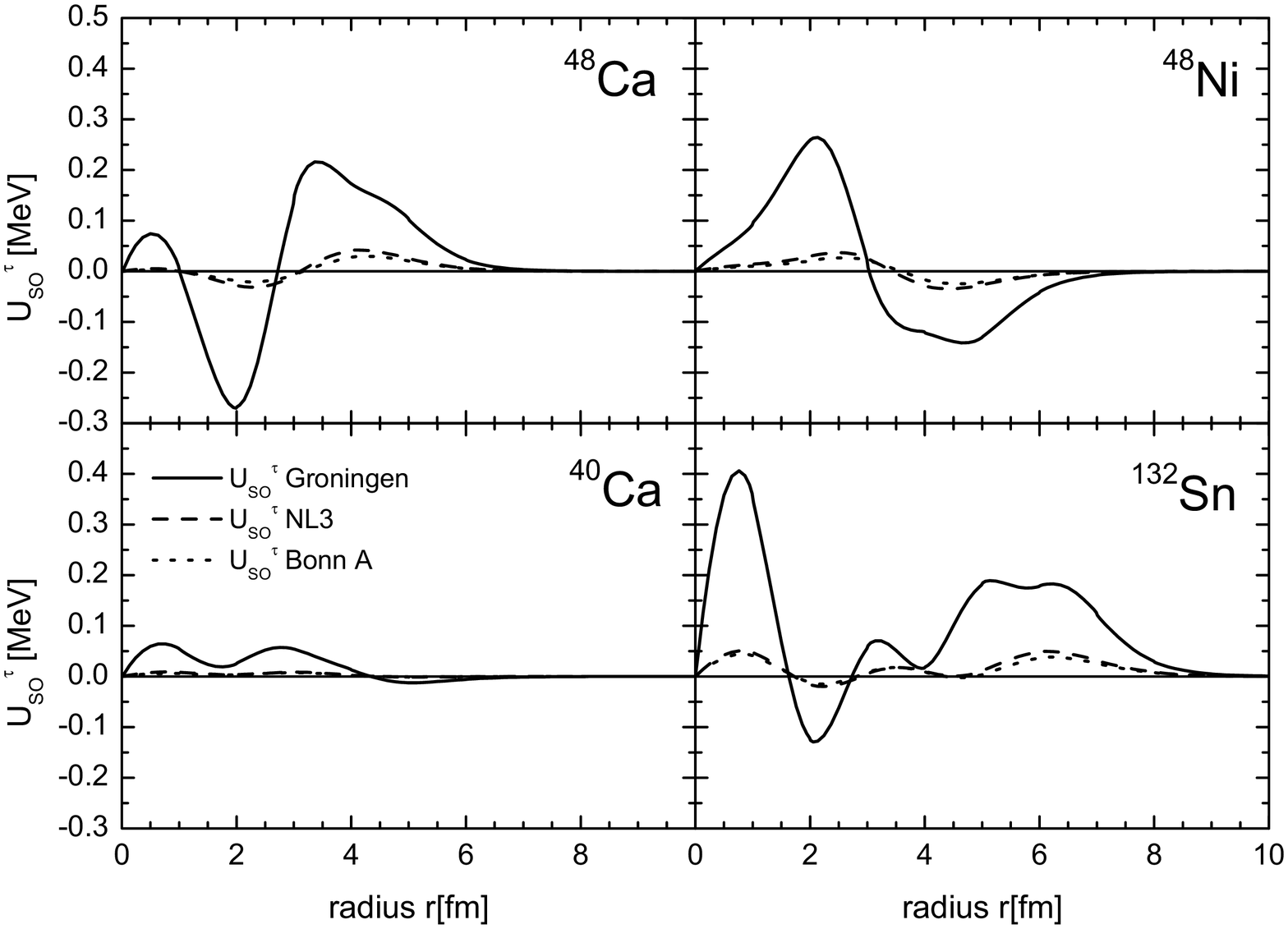,width=8.6cm}
\caption{Same as Fig.~\ref{fig:spin-orbit} but for the isovector spin-
orbit potentials $U_{\tau}^{SO}$.}
\label{fig:spin-orbit_iso}
\end{figure}

Figure \ref{fig:spin-orbit_iso} illustrates the isovector dependence of
the spin-orbit potential discussed in Sec.~\ref{ssec:FiniteProp}.
While $U_{\tau}^{SO}$ has a non-negligible strength in neutron-rich
nuclei and the mirror nuclei $^{48}$Ca and $^{48}$Ni its main
contribution is not located at the nuclear surface. One could expect
to see an effect comparing the spin-orbit splitting of neutrons and
protons or mirror nuclei but we found no systematic behavior. The
isovector spin-orbit potential does not seem to manifest itself
strongly in the single particle energies and is probably overshadowed
by the bulk isovector potential.

\subsection{Ni and Sn isotopes}
\label{ssec:Exotic}

The isotopic chains of Ni and Sn are of particular interest for nuclear
structure calculations because of their proton shell closures at Z=28 (Z=50).
They also extend from the proton dripline that is found nearby the doubly magic
$^{48}$Ni and $^{100}$Sn nuclei to the already $\beta$ unstable neutron-rich
doubly magic $^{78}$Ni and $^{132}$Sn isotopes. This allows us to investigate
the isovector properties of the derived density dependent interactions and to
test the interactions in regions far off stability. An extensive investigation
of these nuclei can be found in \cite{Lalazissis:98,Patyk:99,Mizutori:00} and
the references therein.

\subsubsection{Binding energies}
\label{sssec:BindEner}

\begin{figure}
\centering\epsfig{file=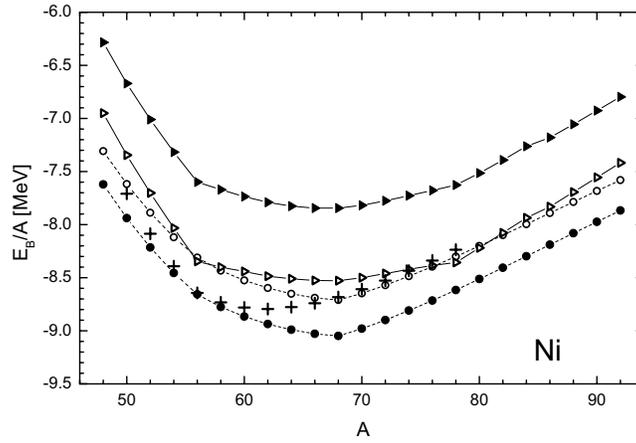,width=8.6cm}
\caption{Binding energies per nucleon for the nickel isotopes. Shown
are results with (open circles) and without momentum correction (solid
circles) for the parameterizations of the Bonn A potential. Results
derived from the Groningen parameter set are denoted by solid
triangles (momentum corrections adjusted to nuclear matter EoS) and
open triangles (adjusted to finite nuclei). Experimental binding
energies are denoted by crosses (taken from 
Ref.~\protect\cite{Audi:93}). 
The lines are drawn to guide the eye.}
\label{fig:BA_Ni}
\end{figure}

\begin{figure}
\centering\epsfig{file=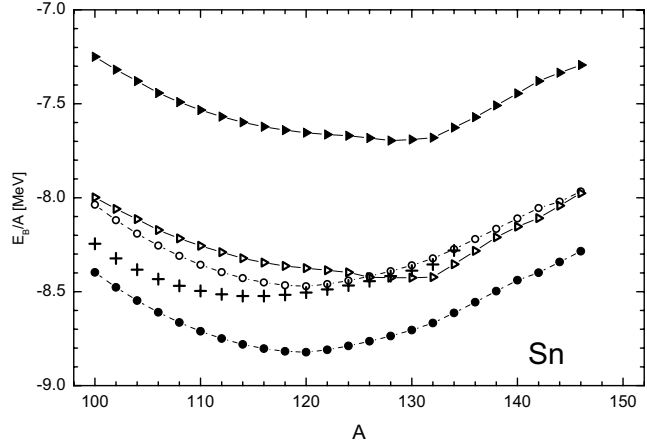,width=8.6cm}
\caption{Same as Fig.~\ref{fig:BA_Ni} but showing binding energies of
the tin isotopes.}
\label{fig:BA_Sn}
\end{figure}

\begin{figure}
\centering\epsfig{file=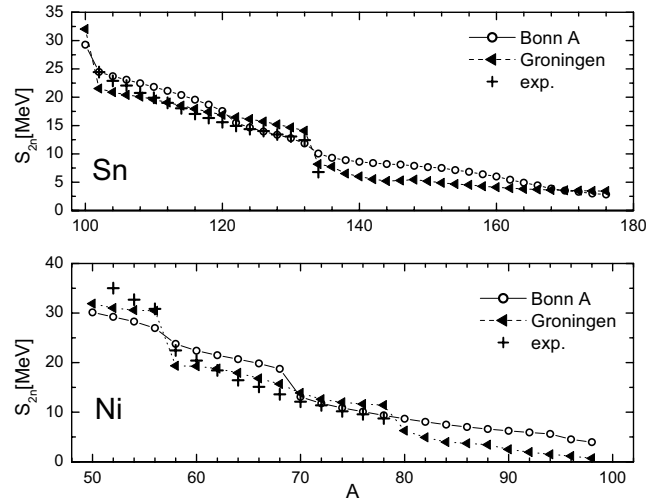,width=8.6cm}
\caption{Two neutron separation energies for the nickel (lower part)
and the tin (upper part) isotopes. Results for the Groningen
(triangles) and Bonn A (circles) parameterization are shown.
Experimental separation energies are denoted by crosses (taken from
Ref.~\protect\cite{Audi:93}). The lines are drawn to guide the eye.}
\label{fig:S2n_Sn-Ni}
\end{figure}

Theoretical and experimental binding energies per nucleon are compared in
Fig.~\ref{fig:BA_Ni} for nickel and in Fig.~\ref{fig:BA_Sn} for tin. The
Groningen interaction derived from nuclear matter strongly underbinds the Ni and
Sn isotopes. The reason for this was already discussed in 
Sec.~\ref{ssec:ClosedShell}. 
The adjusted parameter set describes neutron rich nuclei
reasonably well but fails on the neutron-poor side. Agreement for the Sn
isotopes is relatively poor. This is mainly caused by the extremely strong shell
closure at N=82 that shifts the minimum of the binding energy from the
experimental value of approximately $^{115}$Sn to the doubly magic $^{132}$Sn.
The reason is the too strong spin-orbit splitting of the Groningen potential
that affects the subshell structure of the Sn isotopes and enhances the shell
gap. Nevertheless, the correct shell structure is reproduced as can be seen from
Fig.~\ref{fig:S2n_Sn-Ni} where the two-neutron separation energies
\beq
S_{2n}(Z,N)=B(Z,N)-B(Z,N-2)
\eeq
are compared with experimental values. The agreement is satisfying except for
the too strong shell gaps for $^{132}$Sn and $^{56}$Ni. The momentum correction
does not strongly affect $S_{2n}$ since it mainly shifts the binding energy of
the isotopic chain not affecting the binding energy differences.

The Bonn A parameter set describes the neutron-poor Ni isotopes fairly well but
overbinds the neutron-rich Ni and the Sn isotopes. On the other hand, agreement
of the results for the momentum corrected interaction with experimental data on
the neutron- rich side is very good but binding energies of neutron-poor
isotopes are underestimated by 0.2 MeV. Also, comparing the two-neutron
separation energies (Fig.~\ref{fig:S2n_Sn-Ni}), experimental results are
reproduced well, but some deviations can be noticed. The shell gaps for $^{132}
$Sn and $^{56}$Ni are relatively small whereas the gap for $^{68}$Ni is too
large. This also explains why the minimum of the binding energy is found at
$^{68}$Ni and the isotopes of the $^{56}$Ni to $^{68}$Ni shell are underbound.
The reason for this probably lies in the weak spin-orbit splitting of the Bonn A
potential.

Realizing that for N $\rightarrow$ Z nuclei only the Coulomb interaction
determines the difference between the neutron and proton energies while the
isovector interaction is strongly suppressed we conclude that the underbinding
of the neutron- poor nuclei is caused by a too weak central isoscalar potential.
On the other hand the neutron- rich isotopes are mostly affected by the
isovector interaction that acts attractive for protons, balancing the Coulomb
repulsion, and repulsive for neutrons. Since neutron-rich isotopes are described
well this is an indication that for both the Groningen and the Bonn A potential
the microscopic isovector potentials seem to be too weak. One can visualize this
easily by shifting the curves in Figs.~\ref{fig:BA_Ni} or \ref{fig:BA_Sn}
vertically to reproduce the binding energy of $N=Z$ nuclei. This increases in
first order only the isoscalar central potential. Then, the neutron-rich
isotopes are overbound, indicating that the repulsion from the isovector
potential is too weak.

\subsubsection{Density distributions}
\label{sssec:DensDis}

\begin{figure}
\centering\epsfig{file=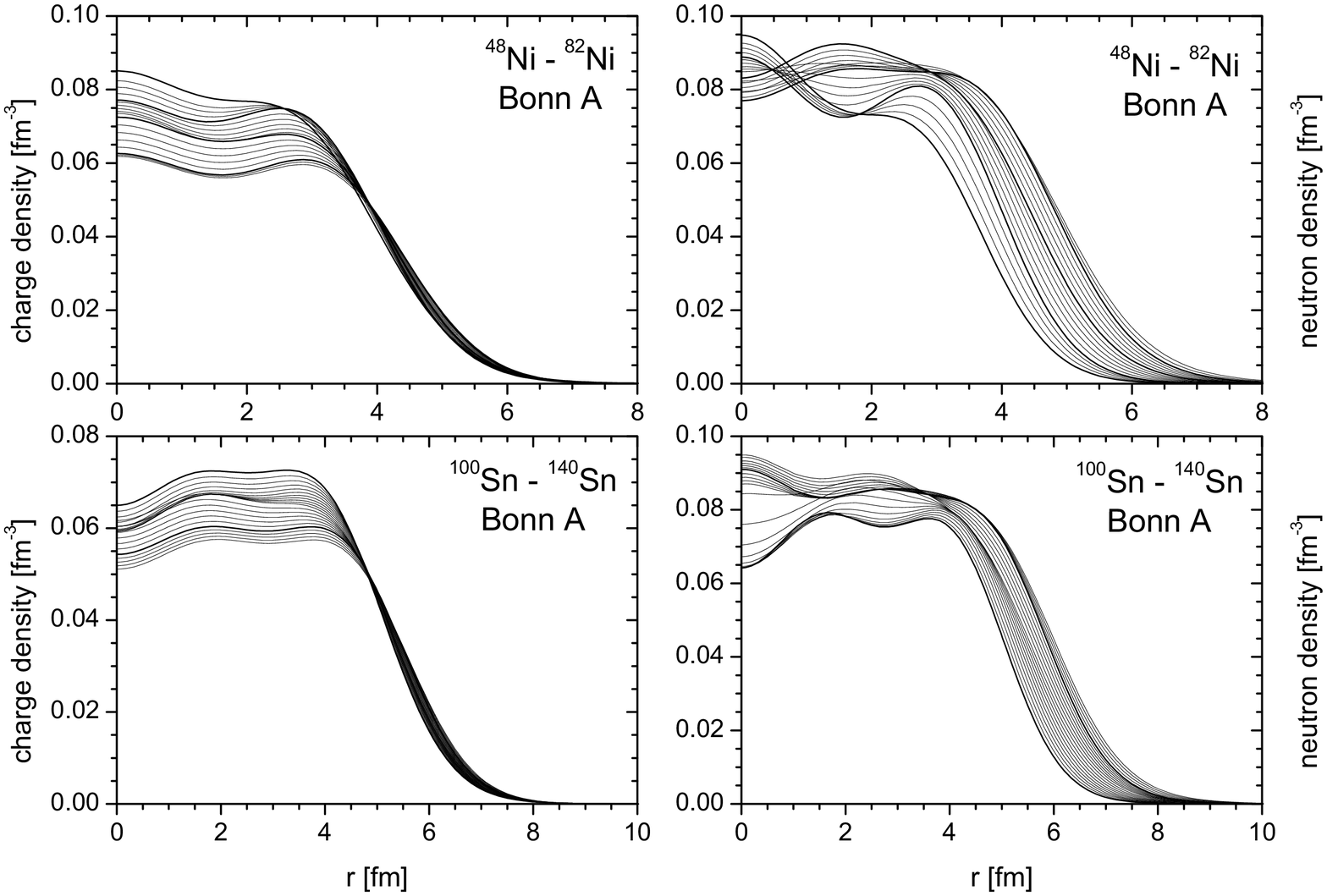,width=8.6cm}
\caption{Charge and neutron density distributions for the isotopic
chains of the Sn and Ni nuclei, calculated with the Bonn A parameter
set.}
\label{fig:densities}
\end{figure}

\begin{figure}
\centering\epsfig{file=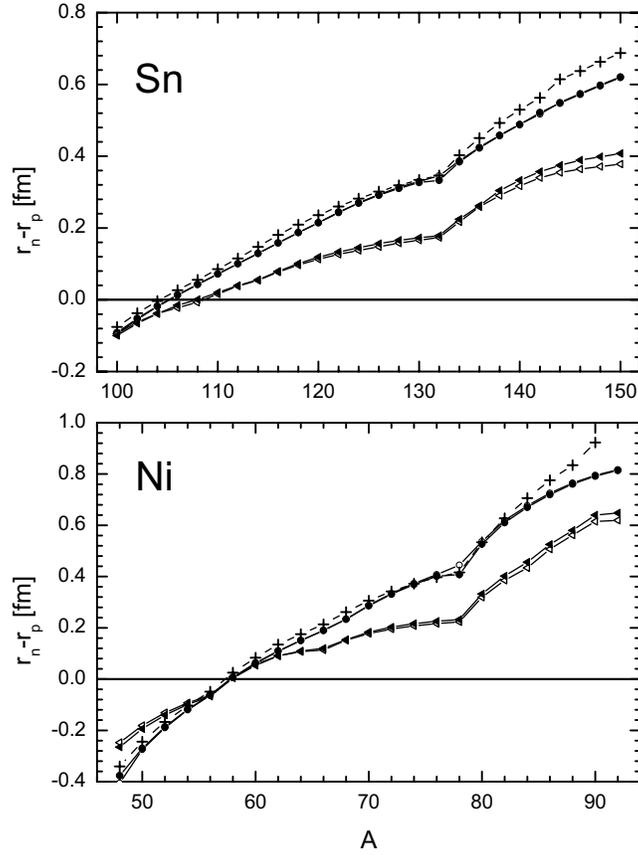,width=8.6cm}
\caption{Difference of the root mean square radii of neutrons and
protons of the nickel (lower part) and the tin (upper part) isotopes
calculated with the Groningen (triangles) and Bonn A (circles)
parameterizations. For comparison results calculated with the
phenomenological NL3 RMF parameter set are shown (crosses). The lines
are drawn to guide the eye.}
\label{fig:rc-rn}
\end{figure}

Self-consistent neutron density distributions and charge densities for the Ni
and Sn isotopes are displayed in Fig.~\ref{fig:densities}. Closed shell nuclei
are marked by bold lines. The distributions are only presented for the Bonn A
potential but for the Groningen potential the same systematics is obtained. For
the calculation of the charge density the theoretical point particle density
distribution $\rho_p$ is folded with a Gaussian proton form factor
\cite{Vautherin:72} with $\sqrt{\langle r^2\rangle_p}=0.8$ fm. From $^{48}$Ni to
$^{72}$Ni the charge density in the interior is reduced by about 30\%
accompanied by a mild increase of the charge radius by about 5\%. For the
neutron densities a more drastic evolution is found. Beyond $^{78}$Ni (N=50
shell closure) the $2d_{5/2}$ subshell is being filled and a thick neutron skin
is build up. This leads to sudden jump in the neutron rms radii while the proton
rms radii increase only slowly. This behavior is more clearly visible in
Fig.~\ref{fig:rc-rn} where the differences of the proton and neutron rms radii
are shown. Approaching the proton dripline a relative thick proton skin below 
N=28 is predicted. 
We also compare our results to calculations with the NL3
interaction. In Ref.~\cite{Lalazissis:98} relativistic Hartree-Bogoliubov
calculations for the Ni and Sn isotopes using this interaction were performed
leading to excellent agreement with experimental data. Figure \ref{fig:rc-rn}
shows very good agreement for the Bonn A potential with the results for the NL3
interaction. The Groningen potential has the same tendency but with smaller
values. This is explained by the systematic underestimation of the rms radii
(Fig.~\ref{fig:groningen_magic}) that also causes a reduced difference between
proton and neutron rms radii. Nevertheless, the same neutron skin is found.
Since the Groningen potential leads to larger shell gaps for reasons of its
strong spin-orbit potential, at the (sub)shell closure N=40 a small increase is
already seen. Figure \ref{fig:rc-rn} also illustrates that the results are
independent of the momentum correction. This is obvious since the correction
only shifts all rms radii to higher or lower values.

The same calculations were performed for the Sn isotopes. In
Fig.~\ref{fig:densities} density distributions from $^{100}$Sn to
$^{140}$Sn are displayed showing the same neutron skin as the Ni
isotopes. Here, one finds a sudden jump beyond N=82 where the
$1h_{11/2}$ shell is filled and the $2f_{7/2}$ subshell becomes
populated (Fig.~\ref{fig:rc-rn}). Again the agreement between the Bonn
A and the NL3 parameter set is excellent. In Ref.~\cite{Hofmann:98} 
calculations for the Sn isotopes with non-relativistic
interactions were performed and identical results were found. This
shows that the observed neutron skins in Sn and Ni are relatively
independent of the NN interaction and the theoretical approach.

\section{Summary and conclusion}
\label{sec:Summary}

We have extracted density dependent meson-nucleon vertices from DBHF self-
energies of asymmetric nuclear matter, derived from realistic NN potentials. For
the Groningen NN potential we found that the coupling constants can be expressed
solely in terms of the nuclear matter baryon density $\rho$ and are independent
of the asymmetry fraction for which the self-energies were calculated. The
extraction of momentum- independent self-energies introduces errors in the
density dependence of the interaction and leads to difficulties to reproduce the
DB EoS in the DDRH approach. We introduced a momentum correction to account for
this error and found excellent agreement of our calculated EoS for all asymmetry
fractions $a_s=\rho_p/\rho$ with the Brueckner results. While the corrections
are very small, the sensitivity of the saturation point of nuclear matter and
the binding energies of finite nuclei to this correction is very high. This
indicates the difficulty of extracting momentum independent coupling constants.

Applying the momentum corrected interaction to finite nuclei, we find that the
agreement with experimental results is satisfactory, taking into account that
the used parameterization contains no free parameters for finite nuclei. For the
Groningen parameter set, the binding energy per nucleon is underestimated by 0.5
to 1 MeV whereas the charge radii are about 0.1 to 0.15 fm too small. These
results are comparable with other mean-field calculations 
\cite{Ineichen:95,Shen:97,Cescato:98}. 
Using the Bonn A potential results are improved, indicating
that the Groningen potential is not attractive enough, especially in the low
density regime. We also adjusted the momentum correction factors to improve the
description of the properties of finite nuclei. This leads to a good agreement
with experimental data while keeping the excellent reproduction of the DB self-
energies.

The inclusion of the $\delta$ meson introduces different effective masses for
protons and neutrons and strongly enhances the isovector spin-orbit potential.
However, we found no systematic effect in the isovector spin-orbit splitting. In
order to further examine this effect, detailed experimental data for the single-
particle energy levels and for exotic nuclei are necessary. Compared to
phenomenological RMF interactions, we found the spin-orbit splitting of the
Groningen potential to be enhanced and the one of the Bonn A potential to be
suppressed. This becomes visible in the shell structure of some exotic nuclei.
We also found indications that the isovector interaction of both microscopic
interactions seems to be too weak to reproduce the complete isotopic chains of
Sn and Ni correctly.

In general, we think that the results of the DDRH approach are quite
satisfactory and that the momentum correction provides a consistent scheme to
reproduce DBHF calculations and improve the agreement with finite nuclei.
Improvements of the results could possibly be achieved by going beyond the
ladder approximation and including, e.g., three-body interactions and ring
diagrams. In future investigations we also plan to apply the density dependent
interactions to neutron stars to gain additional insights in the properties of
the isovector density dependence.

\acknowledgments{This work was supported in part by DFG (Contract No.
Le439/4-3), GSI
Darmstadt, and BMBF.}




\begin{table}
\begin{tabular}{ldddd}
meson $\alpha$   & $\sigma$ & $\omega$ & $\delta$ & $\rho$\\
\hline
$m_{\alpha}$[MeV]& 550      & 783      & 983      & 770      \\ \hline
$a_{\alpha}$     & 13.1334 & 15.1640 & 19.1023 & 12.8373 \\
$b_{\alpha}$     &  0.4258 &  0.3474 &  1.3653 &  2.4822 \\
$c_{\alpha}$     &  0.6578 &  0.5152 &  2.3054 &  5.8681 \\
$d_{\alpha}$     &  0.7914 &  0.5989 &  0.0693 &  0.3671 \\
$e_{\alpha}$     &  0.7914 &  0.5989 &  0.5388 &  0.3598 \\ \hline
\multicolumn{5}{c}{$\rho_0 = 0.16$ [$fm^{-3}$]} \\
\end{tabular}
\centering
\caption{Parameterization of the density dependent couplings from
equation (\ref{eq:DDrat}) extracted from DB calculations in asymmetric
nuclear matter\protect\cite{deJong:98a}.}
\label{tab:coupling}
\end{table}
\begin{table}
\begin{tabular}{ccccc}
$a_{\alpha}$ & $b_{\alpha}$ & $c_{\alpha}$ & $d_{\alpha}$ &
$e_{\alpha}$  \\
19.6270 & 1.7566 & 8.5541 & 0.7783 & 0.5746 \\
\hline
\multicolumn{5}{c}{$\rho_0 = 0.16$ [$fm^{-3}$]} \\
\end{tabular}
\centering
\caption{Parameterization of the $\rho$ meson-nucleon vertex after
adjusting the neutron matter DDRH equation of state to the DB binding
energy.}
\label{tab:RhoAdjusted}
\end{table}
\begin{table}
\begin{tabular}{rccc}
                         & DBHF  &  DDRH & DDRH corr. \\
\hline
$\rho_{sat}$ [$fm^{-3}$] & 0.182 & 0.161  &  0.180 \\
$\epsilon/ \rho$ [MeV]   & -15.5 & -13.13 & -15.60 \\
$K$ [MeV]                & --    & 211    &  282   \\
$m^*/M$                  & --    & 0.592  &  0.554 \\
$a_4$ [MeV]              & 25    & 28.2   &  26.1  \\
\end{tabular}
\centering
\caption{Comparison of infinite nuclear matter properties obtained in
the DDRH model with results from DB calculations for the Groningen NN
potential. Results are shown for calculations with and without
momentum correction.}
\label{tab:NucMatResults}
\end{table}
\begin{table}
\begin{tabular}{crr}
exp.       & $r_c$[fm] & $E/A$[MeV/A] \\
\hline
$^{16}$O   & 2.74 & 7.98 \\
$^{40}$Ca  & 3.48 & 8.55 \\
$^{48}$Ca  & 3.47 & 8.67 \\
$^{90}$Zr  & 4.27 & 8.71 \\
$^{208}$Pb & 5.50 & 7.87 \\
$^{48}$Ni  &  --  & 7.27 \\
$^{56}$Ni  &  --  & 8.64 \\
$^{68}$Ni  &  --  & 8.68 \\
$^{100}$Sn &  --  & 8.26 \\
$^{132}$Sn &  --  & 8.26 \\
\end{tabular}
\centering
\caption{Experimental values of the rms charge radii and binding
energies per
nucleon for (semi)magic nuclei. Data are taken from
Refs.\protect\cite{Brown:98,Audi:93,Chartier:96,Fricke:95}}
\label{tab:experiment}
\end{table}
\begin{table}
\begin{tabular}{cdddddd}
Groningen  & \multicolumn{2}{c}{$\zeta_{\sigma} = 0.0$}
           & \multicolumn{2}{c}{$\zeta_{\sigma} = 0.00804$ fm$^{-2}$}
           & \multicolumn{2}{c}{$\zeta_{\sigma} = 0.008$ fm$^{-2}$} \\
           & \multicolumn{2}{c}{$\zeta_{\omega} = 0.0$}
           & \multicolumn{2}{c}{$\zeta_{\omega} = 0.00103$ fm$^{-2}$}
           & \multicolumn{2}{c}{$\zeta_{\omega} = -0.002$ fm$^{-2}$}
           \\
           & $r_c$  & $E/A$  & $r_c$  & $E/A$  & $r_c$  & $E/A$  \\
\hline
$^{16}$O   & 2.76 & 5.65 & 2.61 & 7.00 & 2.56 & 7.56 \\
$^{40}$Ca  & 3.47 & 5.96 & 3.30 & 7.52 & 3.24 & 8.16 \\
$^{48}$Ca  & 3.48 & 6.09 & 3.31 & 7.79 & 3.25 & 8.49 \\
$^{90}$Zr  & 4.26 & 6.00 & 4.06 & 7.77 & 3.99 & 8.49 \\
$^{208}$Pb & 5.49 & 5.28 & 5.24 & 7.10 & 5.15 & 7.84 \\
$^{48}$Ni  & 3.77 & 4.68 & 3.54 & 6.28 & 3.46 & 6.95 \\
$^{56}$Ni  & 3.73 & 5.78 & 3.53 & 7.60 & 3.46 & 8.35 \\
$^{68}$Ni  & 3.88 & 6.13 & 3.69 & 7.83 & 3.63 & 8.52 \\
$^{100}$Sn & (4.46) & (5.42) & 4.24 & 7.25 & 4.16 & 8.00 \\
$^{132}$Sn & 4.69 & 5.86 & 4.47 & 7.68 & 4.40 & 8.42 \\
\end{tabular}
\centering
\caption{Root-mean-square charge radii $r_c$[fm] and binding energies
per nucleon
$E_B/A$ [MeV/A] of (semi)magic nuclei from density dependent
relativistic Hartree (DDRH)
calculations using the Groningen NN potential. Results for different
momentum correction
factors $\zeta_{\sigma}$ and $\zeta_{\omega}$ (see text) are shown.
The values in parenthesis
correspond to unbound nuclei.}
\label{tab:MagicResultsGro}
\end{table}
\begin{table}
\begin{tabular}{cdddd}
Bonn A     & \multicolumn{2}{c}{$\zeta_{\sigma} = 0.0$}
           & \multicolumn{2}{c}{$\zeta_{\sigma} = -0.0030$ fm$^{-2}$}
           \\
           & \multicolumn{2}{c}{$\zeta_{\omega} = 0.0$}
           & \multicolumn{2}{c}{$\zeta_{\omega} = -0.0015$ fm$^{-2}$}
           \\
           & $r_c$  & $E/A$  & $r_c$  & $E/A$  \\
\hline
$^{16}$O   & 2.75 & 8.58 & 2.79 & 8.30 \\
$^{40}$Ca  & 3.46 & 9.02 & 3.50 & 8.69 \\
$^{48}$Ca  & 3.49 & 8.96 & 3.53 & 8.63 \\
$^{90}$Zr  & 4.26 & 8.98 & 4.31 & 8.63 \\
$^{208}$Pb & 5.53 & 8.17 & 5.60 & 7.82 \\
$^{48}$Ni  & 3.84 & 7.62 & 3.89 & 7.31 \\
$^{56}$Ni  & 3.79 & 8.66 & 3.84 & 8.31 \\
$^{68}$Ni  & 3.88 & 9.05 & 3.93 & 8.71 \\
$^{100}$Sn & 4.51 & 8.40 & 4.57 & 8.04 \\
$^{132}$Sn & 4.74 & 8.67 & 4.78 & 8.32 \\
\end{tabular}
\centering
\caption{Same as Table \ref{tab:MagicResultsGro} but with results for
the Bonn A NN potential.}
\label{tab:MagicResultsBoA}
\end{table}
\begin{table}
\begin{tabular}{lcccc}
$\Delta_{LS}(n,p)$ & $^{16}$O & $^{40}$Ca & $^{48}$Ca & $^{48}$Ni \\
\hline
Gron.      & 8.0/7.8 & 8.3/8.0 & 8.0/7.7 & 7.5/7.4 \\
Gron. adj. & 9.0/8.7 & 9.6/8.8 & 8.8/8.5 & 8.3/8.2 \\
\hline
Bonn A     & 4.2/4.2 & 4.6/4.6 & 4.0/4.1 & 3.9/3.8 \\
Bonn A adj.& 4.0/4.0 & 4.4/4.3 & 3.7/3.8 & 3.6/3.6 \\
\hline
exp.       & 6.1/6.3 & 6.3/7.2 & 5.6/4.3 & -- \\
\end{tabular}
\centering
\caption{Neutron and proton spin-orbit splitting $\Delta_{LS}(n,p)$ in
MeV for the $1p$ shell ($^{16}$O) and the $1d$ ($^{40}$Ca, $^{48}$Ca
and $^{48}$Ni) shell. Results are shown for the Groningen and the Bonn
A NN potential, for the interaction derived from nuclear matter and
for the adjustment to finite nuclei, respectively. For experimental
values see Ref.\protect\cite{Fuchs:95}}
\label{tab:spin-orbit}
\end{table}
\end{document}